\documentclass[aps,preprint,superscriptaddress,nofootinbib,preprintnumbers]{revtex4-1}
\pdfoutput=1
\usepackage[switch]{lineno}

\def\aeq{&=&}

\usepackage{amsmath,amssymb,amsfonts}
\usepackage{bm, slashed}
\usepackage{lineno}
\usepackage{makecell}
\usepackage[dvipdfmx]{color,graphicx}
\usepackage{subfigure}
\usepackage{multirow} 
\usepackage[colorlinks=true]{hyperref}
\usepackage{xcolor}
\definecolor{red}{rgb}{0.9, 0,0}

\usepackage{hyperref}

\newcommand{\bw}{\begin{widetext}}
\newcommand{\ew}{\end{widetext}}
\newcommand{\be}{\begin{equation}}
\newcommand{\en}{\end{equation}}
\newcommand{\bee}{\begin{equation}}
\newcommand{\ene}{\end{equation}}
\newcommand{\bea}{\begin{eqnarray}}
\newcommand{\ena}{\end{eqnarray}}
\newcommand{\bes}{\begin{subequations}}
\newcommand{\ens}{\end{subequations}}
\newcommand{\bef}{\begin{figure}}
\newcommand{\enf}{\end{figure}}
\newcommand{\eq}[1]{Eq.~(\ref{#1})}

\def\thefootnote{\fnsymbol{footnote}}
\def\ie{{\it i.e.\ }}
\def\etc{{\it etc.\ }}
\def\etal{{\it et al.\ }}
\def\fb{\, {\rm fb}}

\def\pslash{p\!\!\!\slash }
\def\kslash{k\!\!\!\slash }
\def\qslash{q\!\!\!\slash }

\def\Qslash{Q\!\!\!\!\slash }
\def\p{\partial}

\def\calk{\mathcal{K}}
\def\call{\mathcal{L}}
\def\calm{\mathcal{M}}

\def\calo{\mathcal{O}}
\def\calp{\mathcal{P}}

\def\cals{\mathcal{S}}
\def\calt{\mathcal{T}}
\def\calu{\mathcal{U}}

\def\to{\rightarrow}

\def\tev{{\rm TeV}}
\def\gev{{\rm GeV}}
\def\pslash{p\!\!\!\slash }

\def\ittb{t\bar{t}}

\def\eeb{e^{-}e^{+}}
\def\ihe{\lambda_{V}}

\def\ihtm{\lambda_{\psi_{t}}}

\def\eh{\sigma_{e}}
\def\ebh{\sigma_{\bar{e}}}
\def\l{\ell} 
\def\lb{\bar{\ell}}

\def\t{t} 
\def\tb{\bar{t}}
\def\th{\sigma_{t}} 
\def\tbh{\sigma_{\bar{t}} }

\def\topn{\psi_{\t}}

\begin{document}


\title{Probing CP violation in $e^{+}e^{-}$ production of the Higgs
boson and toponia\vspace{1cm}}

\author{Kaoru Hagiwara}
\email[Electronic address: ]{kaoru.hagiwara@kek.jp}
\affiliation{KEK Theory Center, 1-1 Oho, Tsukuba, Ibaraki 305-0801, Japan}
\affiliation{Sokendai, 1-1 Oho, Tsukuba, Ibaraki 305-0801, Japan}
\affiliation{Department of Physics, University of Wisconsin-Madison,
Madison, WI 53706, USA} 

\author{Kai Ma}
\email[Electronic address: ]{makainca@yeah.net}
\affiliation{KEK Theory Center, 1-1 Oho, Tsukuba, Ibaraki 305-0801, Japan}
\affiliation{Sokendai, 1-1 Oho, Tsukuba, Ibaraki 305-0801, Japan}
\affiliation{School of Physics Science, Shaanxi University of
Technology, Hanzhong 723000, Shaanxi, China} 

\author{Hiroshi Yokoya}
\email[Electronic address: ]{hyokoya@kias.re.kr}
\affiliation{KEK Theory Center, 1-1 Oho, Tsukuba, Ibaraki 305-0801, Japan}
\affiliation{Quantum Universe Center, KIAS, Seoul 130-722, Republic of
Korea\vspace{2cm}} 
 
\date{\today}

\begin{abstract}
\vspace{0.5cm}
 We study the CP violation in the Higgs boson and toponia production
 process at the ILC where the toponia are produced near the threshold.
 With the approximation that the production vertex of the Higgs boson
 and toponia is contact, and neglecting the P-wave toponia, we
 analytically calculated the density matrix for the production and decay
 of the toponia.
 Under these assumptions, the production spectrum of the toponia is
 solely determined by the spin quantum number, therefore the toponia can
 be either singlet or triplet. 
 We find that the production rate of the singlet toponium is highly
 suppressed, and behaves just like the production of a P-wave
 toponia.
 In the case of the triplet toponium, three completely independent CP
 observables, namely azimuthal angles of lepton and anti-lepton in the
 toponium rest-frame as well as their sum, are predicted based on our
 analytical results, and checked by using the tree-level event
 generator.
 The non-trivial correlations come from the longitudinal-transverse
 interferences for the azimuthal angles of leptons, and
 the transverse-transverse interference for their sum.
 These three observables are well defined at the ILC, where the rest
 frame of the toponium can be reconstructed directly.
 Furthermore, the QCD-strong corrections, which are important near the
 threshold region, are also studied with the approximation of
 spin-independent QCD-Coulomb potential. 
 While the total cross section is enhanced, the spin
 correlations predicted in this paper are not affected.
\end{abstract}

\preprint{KEK-TH-1877}

\maketitle

\tableofcontents

%
\setcounter{page}{1}
\renewcommand{\thefootnote}{\arabic{footnote}}
\setcounter{footnote}{0}

\section{Introduction}
Precise measurements of various physical properties of the observed
Higgs particle $h(125)$~\cite{ATLAS:Higgs2012,CMS:Higgs2012} are the
most important and urgent tasks in the elementary particle physics.
Of particular interest is the property under the charge conjugation and 
the parity transformation, which is called the CP property.
In general, the mass eigenstate $h(125)$ can be either CP eigenstate or
a mixture of CP-even and CP-odd scalar particles.
While only one CP-even scalar particle is predicted in the standard
model (SM), many of its extensions not only modify the Higgs couplings
to gauge bosons and fermions, but also predict additional scalar and
pseudo-scalar particles.
Therefore decisive measurement of the CP property of $h(125)$ can tell
directly whether the observed boson is the Higgs boson in the SM, or it
is described by the model beyond the SM.
The CP property of $h(125)$ has been investigated experimentally by both
ATLAS and CMS collaborations~\cite{ATLAS:Higgs:CP2013,CMS:Higgs:CP2013,%
CMS:Higgs:CP4l2014} through the decays into vector boson pairs, and
the experimental results disfavor the pure CP-odd hypothesis by nearly 
$3\sigma$.
However a large CP mixing has not been excluded
yet~\cite{Kobakhidze:2014gqa,Ellis:2013,Abe:2014,Nishiwaki:2014,Klevtsov:2014,Bolognesi:2012, Brod:2013, Shu:2013, Dolan:2014, Chen:2014-1,%
 Chen:2014-2, Bishara:2014, Korchin:2013, Bernreuther:2010,Plehn:2002,Aquila:1989-1, Aquila:1989-2, Harnik:2013,%
Bower:2002, Desch:2003, Berge:2008, Berge:2009, Berge:2011, Berge:2013,Hagiwara:2015}. 
The reasons are twofold: first, the CP-even coupling of Higgs to the $Z$
boson pair appears at the tree level while the CP-odd coupling appears
only at the loop level; second, the branching ratio to the $ZZ^{\star}$
is small.

Theoretically, the CP property of $h(125)$ can also be measured by
studying the spin correlations of the two jets in the $pp\to hjj$
process~\cite{Plehn:2002}, and the spin correlation in the
$h\to\tau\tau$ channel~\cite{Aquila:1989-1, Aquila:1989-2, Harnik:2013,%
Bower:2002, Desch:2003, Berge:2008, Berge:2009, Berge:2011, Berge:2013}.
For the process $pp\to hjj$, the QCD backgrounds can significantly
reduce the signal significance~\cite{Plehn:2002}, even through the
jet-matching technique can be useful to select out the
signals~\cite{Hagiwara:2015}. On the other hand, the CP properties of the Higgs boson can be
investigated by using the Higgs coupling to top quarks which is the
largest Yukawa coupling in the SM, $y_{t}\sim\calo(1)$. The ATLAS group have studied the $t\bar{t}h$ production with an integrated luminosity of $20.3{\rm fb}^{-1}$, and set a $95\%{\rm C.L.}$ limit on the cross section $\sigma_{t\bar{t}h} < 4.1\sigma_{t\bar{t}h}^{SM}$. However, the information on the CP properties of the top-Higgs Yukawa coupling is still lacking. In Ref.~\cite{Kobakhidze:2014gqa}, the authors shown that the CP mixing parameter is limited in the range $\xi_{htt} < 0.6\pi$.  In Refs.~\cite{Ellis:2013,Abe:2014,Nishiwaki:2014} constraints on the CP-odd $htt$ coupling is studied by using the LHC run-I data through the $hgg$ and $h\gamma\gamma$ couplings. These constraints are not strong, and still allowing a wide range of the CP-mixing angles. In Ref.~\cite{Klevtsov:2014}, a strong constraint on the CP-odd $htt$ coupling is derived by using the constraints on electric dipole moments for several nucleus. However, this constraint is obtained under the assumption that the CP-odd $htt$ coupling is only the source of CP violation, which means there is no contribution from heavier Higgs bosons, sparticle, electron-Higgs CP-odd couplings, \etc. If there are other sources of CP violation and there is a cancelation between them, the constraint can be weakened.

As well, there have also appeared many papers devoted to find optimized CP
observables at hadron colliders~\cite{Kolodziej:2015qsa,Buckley:2015,%
Casolino:2015cza,Brod:2013cka,Ellis:2013yxa,Demartin:2014fia,%
Boudjema:2015nda,Nishiwaki:2013cma,He:2014xla} and
lepton colliders~\cite{Bhupal:2008,Godbole:2007,Godbole:2011,%
Ananthanarayan:2014eea,Ananthanarayan:2013cia}. 
The simplest one requires the reconstruction of the top- and
anti-top-quark momenta from their decay products which is difficult to
be measured accurately even at lepton colliders.
In principle, one can construct CP-odd observables by replacing the top-
and anti-top-quark momenta by the momenta of the $b$ and $\bar{b}$-jets
from the $t$ and $\bar{t}$ decays, respectively.
However, the sensitivity to the CP violating effects gets diluted in this
partial reconstruction.
It has also been pointed out that the different phase-space
distributions for scalar and pseudo-scalar Higgs boson production rates can be
used to determine the CP properties of the $\t\tb h$ coupling.
In Ref.~\cite{Bhupal:2008}, the authors have demonstrated that the CP
properties of Higgs can be assessed by measuring just the total cross
section and the top-quark polarization.
However, these two observables are CP-even, hence only proportional
to the square of the CP-odd coupling.
Furthermore, the ratio of the production rates for pseudo-scalar and for
scalar is very small unless $\sqrt{s} \gg 1~\tev$ where the chiral limit
is recovered.
Therefore, the experimental sensitivities of these observables are not as good as enough to probe
small CP-odd coupling.
To pin down the CP property of the Higgs boson, true CP-odd observables,
which is linearly proportional to the CP-odd coupling are really
required. 
The up-down asymmetry of the momentum direction of the anti-top quark
with respect to the top-quark-electron plane is an example of such an
observable~\cite{Godbole:2007,Godbole:2011}. 
However, the asymmetry is due to the interferences between the
amplitudes involving the $\t\tb h$ vertex and those involving the $hZZ$
vertex.
It has been shown that the latter contribution is very small, amounting
to only a few percent for $\sqrt{s}\le 1~\tev$~\cite{Bhupal:2008}.
Therefore only about 5\% asymmetry can be
observed at the largest~\cite{Godbole:2007,Godbole:2011}.

In this paper, we study the density matrix for the $e^+ e^-$ production
of the Higgs boson and toponia analytically, and propose new CP-odd
observables for the measurement of the CP property of the Higgs boson at
the ILC with $\sqrt{s}=500~\gev$\cite{Ananthanarayan:2013cia,Ananthanarayan:2014eea,ILC:scenarios,ILC1,ILC2}.
In this energy region, the strong-interaction Coulomb force is known to
be important to calculate the total production rate.
Because the P-wave $\ittb$\footnote{%
Below we call this system universally ``toponium'', no matter if
the real bound state is formed or not.}
production is heavily suppressed, we focus on the S-wave toponia
production.
The contents of this paper is organized as follows.
In Sec.~\ref{sec:effective-vertex} we discuss the effective $\ittb$
production vertex and the spectrum of the toponia in the $e^+e^-\to
t\bar th$ process.
In Sec.~\ref{sec:hela} we present the helicity amplitudes for the
S-wave toponia productions and their decays.
In Sec.~\ref{sec:qcd} we study the QCD bound-state effects for the
$t\bar t$ system.
In Sec.~\ref{sec:nresults} we give the numerical results based on the
tree-level event generator, and discuss the CP asymmetries from leptonic
observables.
Finally, discussions and conclusions are given in
Sec.~\ref{sec:conclusion}.

\section{Effective $t-\bar{t}-h$ vertex}\label{sec:effective-vertex}

In this section we study how the $t\bar{t}h$ interactions affect the
$\t\tb$ system production near the threshold.
We assume that the observed Higgs particle $h(125)$ is a mixture of
CP-even ($H$) and CP-odd ($A$) particles,
\bea\label{eq:mixing}
h \aeq H\cos\xi  + A\sin\xi  \,,
\ena
where $\xi$ is the Higgs mixing angle which is assumed to be real.
For simplicity, we further assume that the Yukawa interactions are CP
conserving, 
\bea
\call_{\rm int.} \aeq - g_{Hff}\bar{\psi}_{f}\psi_{f} H -  i
g_{Aff}\bar{\psi}_{f}\gamma^{5}\psi_{f} A \,, 
\ena
such that the source of CP violation is only in the Higgs mixing angle
$\xi$ in \eq{eq:mixing}.
The interactions between the mass eigenstate $h(125)$ and the fermion
anti-fermion pair are then described by
\bea
\call_{\rm int.} 
\aeq - g_{hff} h \big( \bar{\psi}_{f}\psi_{f} + i
\tan\xi_{hff}\bar{\psi}_{f}\gamma^{5}\psi_{f} \big) \,, 
\ena 
where
\bea
g_{hff} \aeq g_{Hff}\cos\xi \,,
~~~~
\tan\xi_{hff} = \frac{g_{Aff}}{ g_{Hff} }\tan\xi \,.
\ena
It is worth noting that the effective strengths of the CP-violating
$hff$ couplings can be different for each fermion, even if the origin of
CP-violation is only in the mixing parameter $\xi$.
In this paper, we focus on the $h\t\tb$ coupling, and for convenience we
use the symbol $g_{h}$ to denote the overall coupling constant
$g_{h\t\tb}$, \ie $g_{h}=g_{h\t\tb}$.
The assumption of the real mixing parameter is valid when CP violation
in the Higgs sector is mediated mainly by the interactions with new heavy
particles.

\begin{figure}[htb]
\begin{center}
\includegraphics[scale=0.7]{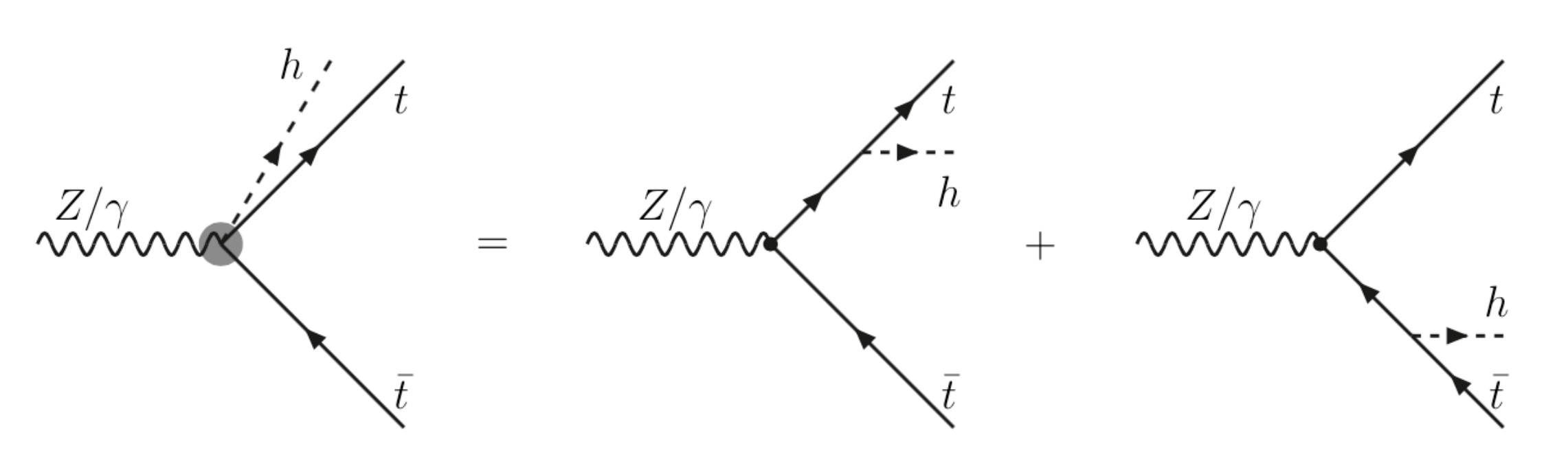}
\end{center}
 \caption{Feynman diagrams which contribute to the $B-h\ittb$ effective
 vertex (labeled by a big gray dot) in the threshold region.
}
\label{Fig:FeynDia:TPP}
\end{figure}

For the $s$-channel production of $t\bar{t}$ associated with $h(125)$, 
\bea
e^{-}(k_{1}, \eh) + e^{+}(k_{2}, \ebh)  &\to&
\t ( p_1, \th ) + \tb ( p_2, \tbh ) + h( k),
\ena
$h(125)$ can be emitted from either a very virtual top-quark or anti-top
quark as shown in Fig.~\ref{Fig:FeynDia:TPP}. 
Even through the Higgs boson can also be produced through the $hBB'$
vertexes ($B= Z, \gamma$), the contributions are negligible (a few
percent for $\sqrt{s}\le 1~\tev$~\cite{Bhupal:2008}) because of the far
off-shell propagator of the vector bosons.
In principle, CP violation can also appear in these vetrices.
However CP violating operators are induced at the one-loop level, and
hence hugely suppressed compared to the CP-even operators. 
Therefore, we do not consider them to simplify the vertex function in
this section.

\begin{figure}[htb]
\begin{center}
\includegraphics[scale=0.6]{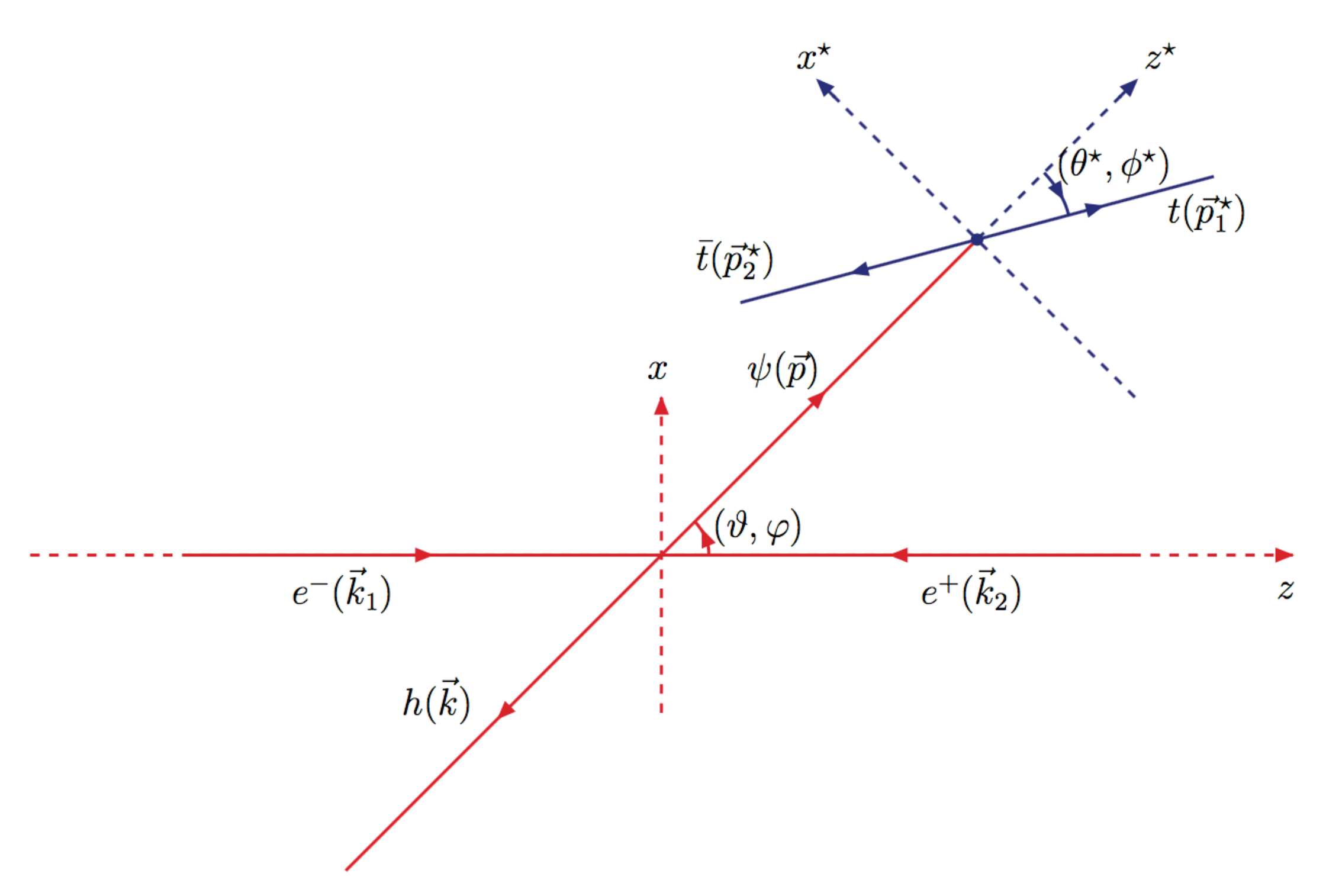}
\caption{Definitions of the kinematical variables in the $e^+e^-$ rest
 frame specified by the axes $x$-$y$-$z$, and the $\ittb$ rest-frame
 specified by the axes $x^{\star}$-$y^{\star}$-$z^{\star}$.
 In the $e^+e^-$ rest-frame, the electron momentum is chosen along the
 $z$-axis and the $\t\tb$ momentum lies in the $x$-$z$ plane with
 positive $x$-component.
 In the $\t\tb$ rest-frame, the $h$ momentum direction is chosen as the
 opposite of the $z^{\star}$-axis, and the $y^{\star}$-axis is taken as
 the same direction as the $y$-axis.}
 \label{fig:kin:production}
\end{center}
\end{figure}

Near the production threshold at $\sqrt{s_{\rm thr.}} = 2m_{t} + m_{h}
\simeq 471~\gev$, the $t\bar{t}h$ system is non-relativistic.
According to the uncertainty principle, the virtual top and anti-top
quarks can propagate only in a distance $\sim 1/(\sqrt{s}-m_{t})$, which
is considerably shorter than the Coulomb radius $r_{C} \sim
1/(\alpha_{s}m_{t})$, at which the QCD interactions bound top and
anti-top quarks to form the bound states toponia.
Therefore treating the whole production vertex as a local interaction
should be a good approximation near the threshold. 
By denoting the vertex of $\t\tb$ production from a virtual vector boson
$B$ ($B=\gamma, Z$) as $\Gamma_{B}^{\mu} = g_{V}^{B\t\tb}\gamma^{\mu} +
g_{A}^{B\t\tb}\gamma^{\mu}\gamma^{5}$, the leading order effective Higgs
radiation vertex is given as 
\bea\label{eq:genEffectiveVertex}
V^{\mu}( p_{1}, p_{2} ) 
\aeq \frac{1}{Q^2 - 2Q\cdot p_{2} }\Gamma_{h}(\Qslash - \pslash_{2} +
m_t)\Gamma^{\mu}_{B} 
- \frac{1}{Q^2 - 2Q\cdot p_{1} }\Gamma^{\mu}_{B}(\Qslash - \pslash_{1} -
m_t) \Gamma_{h},
\ena
where $\Gamma_{h}$ is the abbreviation of the $t\bar{t}h$ vertex which
is $\Gamma_{h} = g_{h}$ for the pure scalar case and $\Gamma_{h}
= g_{h}\tan\xi_{h\t\tb}\gamma_{5}$ for the pure pseudo-scalar case, and
the kinematical variables are defined as in
Fig.~\ref{fig:kin:production} with $Q=k_{1}+k_{2}=p_{1}+p_{2}+k$.
Because both $t\bar{t}$ and $h(125)$ are non-relativistic, the 3-momenta 
$\vec{p}_{1,2}$ could be neglected in the denominators \ie
$p_{1,2}^{\mu} \approx (m_{t},\vec{0})$.
Then the two radiation channels can be combined into a compact form.
For convenience, we expand the spinor structure of this vertex by using
the Clifford algebra as follows:
\bea\label{eq:effvertex}
V^{\mu}( p_{1}, p_{2} ) 
= 
\frac{1}{ s -  2 m_{t}\sqrt{s}  }
\bigg(
c^{\mu}_{S} + c^{\mu}_{P} \gamma^{5} + c^{\mu\nu}_{V}\gamma_{\nu}
+c^{\mu\nu}_{A}\gamma_{\nu} \gamma^{5} 
+\frac{1}{2}c^{\mu\alpha\beta}_{T}\sigma_{\alpha\beta}
\bigg)\,,
\ena
where we have used $Q^2=s$.
The expansion coefficients can be calculated easily, as shown in
Table~\ref{table:CA}. 
The production dynamics are described completely by the vertex function
$V^{\mu}( p_{1}, p_{2} )$ in \eq{eq:effvertex}.
Note that the coefficients of the (CP-even) $hBB'$ vertexes are not
included in Table~\ref{table:CA} for the clarity and compactness of the
table. 
These contributions are very small, a few percent for $\sqrt{s}\le
1~\tev$~\cite{Bhupal:2008}), and can be easily counted by modifying the
coefficients $c_{V}^{\mu\nu}$ and $c_{A}^{\mu\nu}$.
Furthermore, the spin correlation which can be used to measure
the CP violation effects does not depend on the coefficients of these
operators.
The magnitudes of these contributions are discussed in the numerical
simulation part in Sec.~\ref{sec:nresults}.

\begin{table}[htbp]
\renewcommand\arraystretch{1.5}
\caption{The Clifford expansion coefficients in \eq{eq:effvertex}.
 The $B\t\tb$ ($B=\gamma, Z$) vertex is denoted as $\Gamma_{B}^{\mu} =
 g_{V}^{B\t\tb}\gamma^{\mu} + g_{A}^{B\t\tb}\gamma^{\mu}\gamma^{5}$.
 The $h\t\tb$ vertex is denoted as $\Gamma_{h} = g_{h} + i
 g_{h}\tan\xi_{h\t\tb}\gamma_{5}$.
 The momentum $q^{\mu} = p_{1}^{\mu}-p_{2}^{\mu}$ is the relative
 momentum between the top and anti-top quarks.
 Note that the coefficients of the (CP-even) $hBB'$ vertexes are not
 included for the clarity and compactness of the table.}
\label{table:CA}
\begin{center}
\begin{tabular}{c|c|c}
\hline\hline
$\mathcal{O}_{X}$ & {\rm Scalar }($\Gamma_{h}=g_{h}$) & {\rm
	 Pseudo-Scalar }($\Gamma_{h} = g_{h}\tan\xi_{h\t\tb}\gamma_{5}$)
\\[1mm]\hline
$c^{\mu}_{S}$  &  $ g_{h}\, g_{V}^{B\t\tb}\, q^{\mu} $ 
&  
$i \,g_{h}\tan\xi_{h\t\tb}\,g_{A}^{B\t\tb}(Q^{\mu} + k^{\mu})$
\\\hline
$ c^{\mu}_{P} $ & $ g_{h}\, g_{A}^{B\t\tb} \,(Q^{\mu} + k^{\mu}) $ 
& 
$ i\,g_{h}\tan\xi_{h\t\tb}\, g_{V}^{B\t\tb} q^{\mu}$
\\\hline
$ c^{\mu\nu}_{V} $ & $ 2 m_{t}\,g_{h}\, g_{V}^{B\t\tb}\,g^{\mu\nu}$ & $0$
\\\hline
$c^{\mu\nu}_{A}$ & $ 2 m_{t}\,g_{h} \,g_{A}^{B\t\tb}\,g^{\mu\nu}$ & $0$ 
\\\hline
$c_{T}^{\mu\alpha\beta}$ 
& 
\begin{tabular}{c}
$ i g_{h} \,g_{V}^{B\t\tb}\, \big[(Q^{\beta}+k^{\beta})g^{\mu\alpha}-
 (Q^{\alpha}+k^{\alpha})g^{\mu\beta}\big] $;
\\[2mm]
$ g_{h} \,g_{A}^{B\t\tb}\, \epsilon^{\alpha\beta\mu\nu}\, q_{\nu}$
\end{tabular}
& 
\begin{tabular}{c}
$ i \,g_{h}\tan\xi_{h\t\tb} \,g_{V}^{B\t\tb}\,
 \epsilon^{\alpha\beta\mu\nu} (Q_{\nu}+k_{\nu})$  ;
\\[2mm]
$ g_{h} \tan\xi_{h\t\tb}\,g_{A}^{B\t\tb}\, ( q^{\alpha}g^{\mu\beta} -
 q^{\beta}g^{\mu\alpha} ) $
\end{tabular}
\\\hline\hline
\end{tabular}
\end{center}
\end{table}

\begin{figure}[htb]
\begin{center}
\includegraphics[scale=0.6]{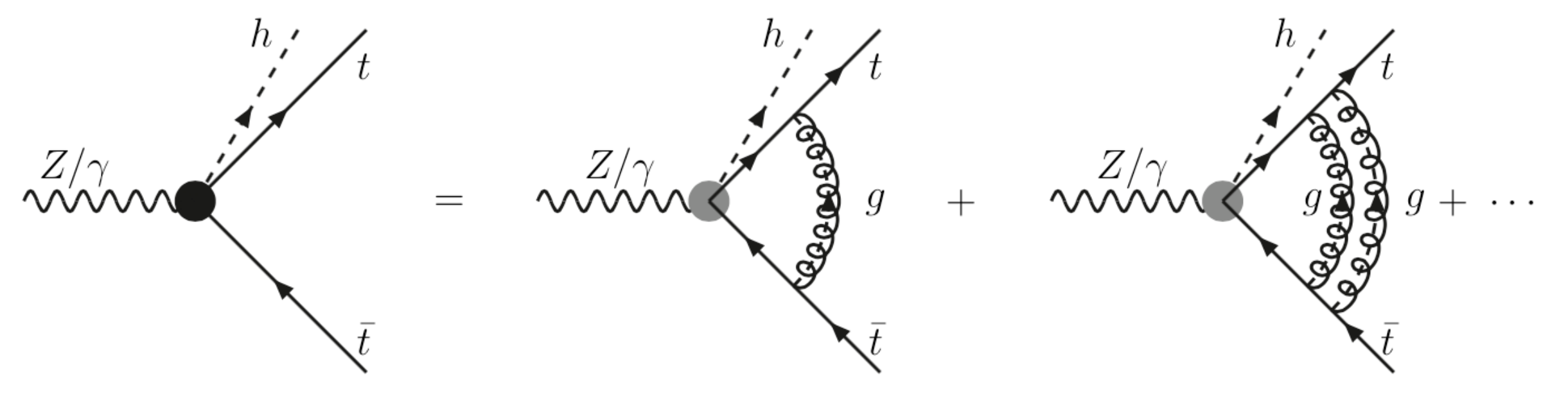}
\end{center}
 \caption{QCD corrections to the effective $V-h\ittb$ vertex in the
 threshold region.
 The big black dot indicates the full vertex function after this
 summation.}
 \label{Fig:sum}
\end{figure}

After the electroweak production of $t\bar{t}h$, the strong interaction
between $t\bar{t}$ becomes important.
In the threshold region, infinite number of Feynman diagrams whose
effects are proportional to the powers of $\alpha_{s}/\beta_{t} \sim
\calo(1)$ contribute, and their resummation is needed; see
Fig.~\ref{Fig:sum}.

After the resummation, the vertex function satisfies an integral
equation, the Salpeter-Bethe equation~\cite{SB:1951}, which describes
the formation of bound states in this region.
We will discuss it carefully in Sec.~\ref{sec:qcd}.
Here we would like to classify the possible bound states that can be
produced.

Table~\ref{table:spectrum} lists the possible bound states up to
P-wave in the spectrum notation for various bi-spinor combinations of
spinors $\psi$ and $\varphi$ (see App.~\ref{app:dirac} for our
conventions of the spinor wave functions in the Dirac representation),
and the corresponding spinor vertex structures in the non-relativistic
limit.
The spin-singlet state can be produced only by the pseudo-scalar
operator $\calo_{P}$ and the time component of the axial-vector operator
$\calo_{V}^{\mu=0}$.
All the other operators can generate the spin-triplet state but with
different angular momentum.

\begin{table}\label{table:spectrum}
\renewcommand\arraystretch{1.5}
\begin{center}
\caption{Quantum numbers of the bi-spinors of top and anti-top quarks in
 the non-relativistic limit in the rest frame of $\t\tb$.}
\begin{tabular}{c|c|c}
\hline\hline
{\rm Operators} 
& 
{\rm Non-relativistic limit }
& 
{\rm  Quantum state} 
\\[1mm]\hline
$\calo_{S}=\bar{\psi}\varphi $ 
& 
$\xi^{\dag} \vec{q}\cdot \vec{\sigma} \eta$ 
& 
$^{3}P_{0}$
\\\hline
$\calo_{P}=\bar{\psi}\gamma^{5}\varphi $ 
& 
$\xi^{\dag} \eta$ 
& 
$^{1}S_{0}$
\\\hline
$ \calo_{V}=\bar{\psi}\gamma^{\mu}\varphi $ 
& 
$(0, \xi^{\dag} \vec{\sigma} \eta)$ 
& 
$^{3}S_{1}$
\\\hline
$\calo_{A}=\bar{\psi}\gamma^{\mu}\gamma_{5}\varphi$ 
& 
$(\xi^{\dag} \eta\,, \,\; \xi^{\dag} \vec{q} \times \vec{\sigma} \,\eta )$ 
& 
$(^{1}S_{0}\,, \,\; ^{3}P_{1})$ 
\\\hline
$\calo_{T}=\bar{\psi}\sigma^{0i}\varphi$ 
& 
$\xi^{\dag} \sigma^{i}\,\eta$ 
& 
$^{3}S_{1}$ 
\\\hline
$\calo_{T}=\bar{\psi}\sigma^{ij}\varphi $ 
& 
$q^{i}\xi^{\dag} \sigma^{j}\,\eta - q^{j}\xi^{\dag} \sigma^{i}\,\eta $
& 
$^{3}P_{1}$
\\\hline\hline
 \end{tabular}
\label{table:spectrum:bispinor}
\end{center}
\end{table}

It should be noted that, all those quantum numbers listed in
Table~\ref{table:spectrum} are also affected by the corresponding
expansion coefficients, which are tabled in Table~\ref{table:CA}.
In Table~\ref{table:spectrum:total}, we show the possible bound states
by combining the coefficients and operators.
For the scalar operator $\calo_{S}$, both the coefficient and bi-spinor
are of P-wave for the scalar Higgs boson.
Therefore the $\t\tb$ system is D-wave which can be ignored completely.
In the case of the pseudo-scalar Higgs boson, the $\t\tb$ system is
P-wave because the coefficient is S-wave.
However, it is still negligible near the threshold region.
For the pseudo-scalar operator $\calo_{P}$, a singlet toponium can be
produced.
The coefficient is S-wave for the scalar Higgs boson, while P-wave for
the pseudo-scalar Higgs boson.
For the vector and axial-vector operators, $\calo_{V}$ and $\calo_{A}$,
only vertexes for scalar Higgs boson production exist.
The operator $\calo_{V}$ can generate the S-wave triplet toponium, while
$\calo_{A}$ generates the P-wave triplet toponium. 
In addition, the axial-vector operator can also generate the S-wave
singlet toponium via its time-component.
This contribution turns out to be very important, because it is
destructive with the contribution of the pseudo-scalar vertex
$\calo_{P}$, and then makes the total production rate of the singlet
toponium highly suppressed.
Of particular interest is the production by the tensor operator
$\calo_{T}$, in which both the bi-spinor and the coefficient contain
S-wave and P-wave $\t\tb$.
Here we discuss only the S-wave contributions.
For both scalar and pseudo-scalar Higgs bosons, it is the ``electric
component'' of the tensor operator $\propto \sigma^{0i}$ generating the
S-wave toponium.

\begin{table}
\renewcommand\arraystretch{1.5}
\begin{center}
 \caption{Quantum states of the $t\bar t$ and $t\bar t h$ systems.
 The spin and angular momenta are summed first by combining the top and
 anti-top-quarks system, and then by combining the toponium ($\topn$)
 and Higgs system.}
\label{table:spectrum:total}
\begin{tabular}{c|c|c|c|c}
\hline\hline
\multirow{2}{*}{\rm Operators} 
& 
\multicolumn{2}{c|}{\rm Scalar Higgs }
& 
\multicolumn{2}{c}{\rm  Pseudo-Scalar Higgs} 
\\
\cline{2-5} & 
$(\t,\tb)$-System  & 
$(\topn,h)$-System & 
$(\t,\tb)$-System  & 
$(\topn,h)$-System 
\\[1mm]\hline
$\calo_{S} $  
&  
$^{3}D_{1}$
&  
$^{3}S_{1}$
&  
$^{3}P_{0}$
&  
$^{1}P_{1}$
\\[1mm]\hline
$\calo_{P} $  
&  
$^{1}S_{0}$
&  
$^{1}P_{1}$
&  
$^{1}P_{1}$
&  
$^{3}S_{1}$
\\[1mm]\hline
$\calo_{V} $  
&  
$^{3}S_{1}$
&  
$^{3}S_{1}$
&  
$0$
&  
$0$
\\[1mm]\hline
\multirow{2}{*}{$\calo_{A} $}  
&  
$^{1}S_{0}$
&  
$^{1}P_{1}$
&  
$0$
&  
$0$
\\\cline{2-5}&  
$^{3}P_{1}$
&  
$^{3}S_{1}$
&  
$0$
&  
$0$
\\[1mm]\hline
\multirow{4}{*}{$\calo_{T} $}  
&  
$^{3}S_{1}$
&  
$^{3}S_{1}$
&  
$^{3}S_{1}$
&  
$^{3}P_{1}$
\\\cline{2-5}&  
$^{3}P_{1}$
&  
$^{3}P_{1}$
&  
$^{3}P_{1}$
&  
$^{3}S_{1}$
\\\cline{2-5}& 
$^{3}P_{1}$
&  
$^{3}S_{1}$
&  
$^{3}P_{1}$
&  
$^{3}P_{1}$
\\\cline{2-5}&  
$^{3}D_{1}$
&  
$^{3}S_{1}$
&  
$^{3}D_{1}$
&  
$^{3}S_{1}$
\\\hline\hline
\end{tabular}
\end{center}
\end{table}

\section{Helicity amplitudes}\label{sec:hela}

In this section we give a formula for the full helicity amplitudes in
terms of the toponium angular momentum.
Near the threshold the QCD-strong interactions become important.
Here we assume the QCD corrections can be completely factorized out,
\ie the strong force is spin-independent; see Sec.~\ref{sec:qcd}.
In this approximation the full physics could be modeled by using pure
electroweak $h\ittb$ production and their decays.
Then, the toponium helicity is obtained by the spin projection.
The spin projection becomes simple when the relative momentum $q^{\mu}$
between the top and anti-top quarks is neglected.
Furthermore neglecting the relative momentum does not lose essential
physics as the top and anti-top quarks have large decay width.
Therefore, while we calculate the density matrix without the assumption
of $|q^{\mu}|\approx 0$, some important results can be discussed under
this simplification.
In subsection~\ref{sec:hela:projection} we give our formalism on the
factorization of the QCD correction, as well as that for the spin
projection.
In subsection~\ref{sec:hela:production} and \ref{sec:hela:decay} we give
the helicity amplitudes for the production and decays of toponia.
The total helicity amplitude and the CP-odd observables are discussed in
subsection~\ref{sec:hela:total}.

\subsection{Factorization and projection of the helicity amplitudes}\label{sec:hela:projection} 

The total amplitude for the process $e^{-} + e^{+} \to h + ( \l
\nu_{\lb}\bar{b} )+(\lb \nu_{\l} b)$ can be written in general as
follows:
\bea
\calm 
\aeq \langle ( \l \nu_{\lb}\bar{b} )(\lb \nu_{\l} b) h | \calt |\eeb
\rangle\,.
\ena
We focus on the CP violation effects due to the anomalous interaction
between the toponia and Higgs boson.
This is done by inserting a complete basis of the $\ittb$ resonance
states $\topn$ with quantum number $(J_{\topn}, \ihtm)$, then the total
helicity amplitude can be written as the product of the production and
decay amplitudes of the toponia,\footnote{
Note that the phase space factor of the toponium has been dropped here,
it will be counted in the phase space part.
Here and after we always drop the phase space factor whenever the
amplitudes are expanded by the complete basis.} 
\bea
\calm 
\aeq \sum_{J_{\topn},\ihtm}
\langle ( \l \nu_{\lb}\bar{b} )(\lb \nu_{\l} b) | \calt_{D}
|\topn(J_{\topn}, \ihtm)
\rangle \langle \topn(J_{\topn}, \ihtm) h| \calt_{P} |\eeb \rangle\,.
\ena
However this amplitude cannot be calculated directly in perturbation
theory because the $\ittb$ resonances $\topn$ are composite states.
We therefore expand the helicity amplitudes by using the fundamental
fields $\t$ and $\tb$, and the amplitudes take the following form: 
\bea
\calm \aeq
\sum_{J_{\topn},\ihtm} \sum_{\th',\tbh'} \sum_{\th,\tbh} 
\calm_{\topn}(J_{\topn}, \ihtm; \th',\tbh';\th,\tbh)
\calm_{D}(\th',\tbh')\calm_{P}(\th,\tbh)\,,
\ena
where the production, decay and resonance amplitudes are, respectively, 
\bea
\calm_{P}(\th,\tbh) &=& \langle  \t(\th) \tb(\tbh) h|\calt_{P} |\eeb
\rangle\,, 
\\[3mm]
\calm_{D}(\th',\tbh') &=& \langle ( \l \nu_{\lb}\bar{b} )(\lb \nu_{\l}
b)  | \calt_{D} | \t(\th') \tb(\tbh') \rangle \,, 
\\[3mm]
\calm_{\topn}(J_{\topn}, \ihtm; \th',\tbh'; \th,\tbh) &=&  
\langle  \t(\th') \tb(\tbh')|\calt_{QCD}^{\dag} |\topn(J_{\topn}, \ihtm)
\rangle  
\langle \topn(J_{\topn}, \ihtm) |\calt_{QCD}| \t(\th) \tb(\tbh)
\rangle\,.
\ena
Here both the production and decay processes are electroweak, and the
QCD corrections are accounted for in the resonance amplitudes.
In order to make our discussions more simple and clear, we use the free
$\t\tb$ resonance states $\widetilde{\psi}_{t}(J'_{\psi}, \ihtm')$ to
separate out the spin degrees of freedom.
Then the amplitude for the toponium formation from the top- and
anti-top-quarks can be written as
\bea
&&
\langle \topn(J_{\topn}, \ihtm) |\calt_{QCD}| \t(\th) \tb(\tbh) \rangle
\nonumber\\[2mm]
&=&
\sum_{J'_{\topn}, \ihtm'} 
\langle \topn(J_{\topn}, \ihtm)
|\calt_{QCD}|\widetilde{\psi}_{t}(J'_{\topn}, \ihtm') \rangle 
\langle \widetilde{\psi}_{t}(J'_{\topn}, \ihtm')|
\calo^{J'_{\psi}}_{\ihtm'}|\t(\th) \tb(\tbh) \rangle\,, 
\ena
where we have introduced a pure kinematical operator
$\calo^{J'_{\psi}}_{\ihtm'}$ to account for the spin correlations of
$\t\tb$ to $\widetilde{\psi}_{\t}$.
In general the quantum numbers $(J_{\topn}, \ihtm)$ can be different
from $(J'_{\topn}, \ihtm')$ by QCD corrections, for instance when we
include the spin-orbital interactions, \etc
Here we neglect those spin-dependent corrections, \ie we take
$(J_{\topn}, \ihtm)=(J'_{\topn}, \ihtm')$.
Then the resonance amplitudes can be written as 
\bea
\calm_{\topn}(J_{\topn}, \ihtm; \th',\tbh'; \th,\tbh) 
\aeq (\calp_{\th',\tbh'}^{J_{\topn},\ihtm})^{\dag}
(\calp_{\th,\tbh}^{J_{\topn},\ihtm} ) 
\calk_{J_{\topn}, \ihtm}
\,,
\ena
where the factor $\calk_{J_{\topn}, \ihtm}$ is defined as the squared
renormalization factor which gives the pure QCD corrections, 
\bea
\calk_{J_{\topn}, \ihtm} \aeq
|\langle \psi(J_{\topn}, \ihtm)
|\calt_{QCD}|\widetilde{\topn}(J_{\topn}, \ihtm) \rangle|^2\,, 
\ena
and the spin projection operator $\calp_{\th,\tbh}^{J_{\psi},\ihtm}$ is
defined as the matrix elements of $\calo^{J_{\topn}}_{\ihtm}$, 
\bea
\calp_{\th,\tbh}^{J_{\psi},\ihtm}  
\aeq \langle \widetilde{\psi}(J_{\psi}, \ihtm)|\calo^{J_{\psi}}_{\ihtm}|
\t(\th) \tb(\tbh) \rangle \,.
\ena
The QCD corrections are discussed in Sec.~\ref{sec:qcd}.
Let us focus on the spin projection first.
In general $J_{\topn}$ can be any integer.
However the production rates of toponium states with higher angular
momentum $L$ are suppressed by $\beta_{t}^{L}$
where $\beta_t$ is a velocity of top and anti-top quarks in the
toponium rest-frame.
Therefore, we discuss only the S-wave resonance.
Then $\topn$ can be either spin-singlet or spin-triplet, \ie
$J_{\topn}=0$ or 1.
The corresponding projection operators are defined as follows:
\bea
\calo^{J_{\topn}=0}_{\ihtm} &=&  \frac{1}{\sqrt{2s_{\topn}} \sqrt{ 1 -
\frac{ (m_{t}^{\star2} -\bar{m}_{t}^{\star2})^2  }{ s_{\topn}^2 } }
}\widetilde{\psi}_{\t} (\ihtm)\bar{t} \gamma^{5} t\,, 
\\
\calo^{J_{\topn}=1}_{\ihtm} &=&  \frac{1}{\sqrt{2s_{\topn}} \sqrt{ 1 -
\frac{ (m_{t}^{\star2} -\bar{m}_{t}^{\star2})^2  }{ s_{\topn}^2 } } }
\widetilde{\psi}_{\t}^{\mu}(\ihtm)\bar{t}\,\gamma_{\mu}t\,, 
\ena 
where $\sqrt{s_{\topn}}$, $m_{t}^{\star}$ and $\bar{m}_{t}^{\star}$ are
the invariant mass of the toponium, top and anti-top quarks
respectively.
The normalization factor is chosen such that the spin projection
operators are dimensionless (the overall normalization of
$\calm_{\topn}$ is fixed by the total QCD correction).
With the help of the spin projection operators the total helicity
amplitude can be expressed in terms of the toponium production and decay
helicity amplitudes as follows:
\bea
\calm 
\aeq  \sum_{J_{\topn},\ihtm} \calk_{J_{\topn}, \ihtm}
\widetilde{\calm}_{P}(J_{\topn},\ihtm)
\widetilde{\calm}_{D}(J_{\topn},\ihtm) \, 
\ena
where the projected production and decay helicity amplitudes are
\bea
\widetilde{\calm}_{P}(J_{\topn},\ihtm) 
&=& 
\sum_{\th,\tbh} \calp_{\th,\tbh}^{J_{\topn},\ihtm}  
\calm_{P}(\th,\tbh) \,,
\\[2mm]
\widetilde{\calm}_{D}(J_{\topn},\ihtm) 
&=& 
\sum_{\th',\tbh'}   (\calp_{\th',\tbh'}^{J_{\topn},\ihtm})^{\dag}
\calm_{D}(\th',\tbh') \,.\label{eq:MD}
\ena
In the next two subsections, we study these two helicity amplitudes.

\subsection{Production helicity amplitudes}\label{sec:hela:production} 

In this subsection we give the helicity amplitudes for the production
process of toponia in associated with the Higgs boson.
The kinematical variables are defined as (see also the
Fig.~\ref{fig:kin:production})
\bea
e^{-}(k_{1}, \eh) + e^{+}(k_{2}, \ebh)  &\to& \widetilde{\topn}( p;
J_{\topn}, \ihtm ) + h( k)  
\to \t ( p_1, \th ) + \tb ( p_2, \tbh ) + h( k).
\ena
The fermion helicities are $\sigma_{i} = \pm1/2$ for $i=e,\bar{e}, \t,
\tb$.
For the spin-singlet toponium $J_{\topn}=0, \ihtm=0$, and for the
spin-triplet toponium $J_{\topn}=1, \ihtm=0, \pm1$.
In the rest frame of $e^{+}e^{-}$ the particle momenta are given by 
\bes\bea
Q^{\mu} &=& \sqrt{s} (1, \,0, \,0, \, 0  )\,,
\\[3mm]
k_{1}^{\mu} &=& \dfrac{ \sqrt{s} }{2} (1, \,0, \,0, \, 1  )\,,
\\[3mm]
k_{2}^{\mu} &=& \dfrac{ \sqrt{s} }{2} (1, \,0, \,0, \, -1  )\,,
\\[3mm]
p^{\mu} &=& \dfrac{ \sqrt{s} }{2} (1+\dfrac{s_{\topn} - m_{h}^2}{s},
\,\beta\sin\vartheta\cos\varphi, \,\beta\sin\vartheta\sin\varphi,
\,\beta\cos\vartheta  )\,, 
\\[3mm]
k^{\mu} &=& \dfrac{ \sqrt{s} }{2} (1-\dfrac{s_{\topn} - m_{h}^2}{s},
\,-\beta\sin\vartheta\cos\varphi, \,-\beta\sin\vartheta\sin\varphi,
\,-\beta\cos\vartheta  )\,.
\ena\ens
Here we use $\sqrt{s}$ to denote the total collision energy, and
$\sqrt{s}_{\topn}$ to denote the invariant mass of the toponium.
$\beta$ is a velocity of the Higgs boson and toponium in this frame,
which is given as 
\bea
 \beta \aeq \sqrt{1+\frac{m_h^4}{s^2}+\frac{s_{\topn}^2}{s^2}
 -\frac{2m_h^2s_{\topn}}{s^2} - \frac{2m_h^2}{s}-\frac{2s_{\topn}}{s}}.
\ena
In this frame the leptonic current is give by
\bea
L_{V}^{\mu}(\lambda_{e}) \aeq -\ihe {\cal G}_{\ihe}^{e}
\sqrt{2s}\,\varepsilon^{\mu}(\vec{Q}=\vec{0}, \ihe)\,, 
\ena
where $\varepsilon^{\mu}(\vec{Q}=\vec{0}, \ihe)$ are given in \eq{eq:vectorWaveT} and \eq{eq:vectorWaveL} in the Appendix \ref{subsec:vectorWave} by setting $\theta=0$ and $\phi=0$, $\ihe = \eh -\ebh = \pm1$ is the helicity of the virtual vector
particle $B$ that can be either photon ($B=\gamma$) or $Z$ ($B=Z$); the
helicity-dependent form-factor ${\cal G}_{\ihe}^{e}$ is defined as
\bea
 {\cal G}_{\ihe}^{e}(Q^2)  \aeq
\begin{cases}
\dfrac{ e }{ Q^2 } & \text{for}\ B=\gamma \,
\\[5mm]
- \dfrac{1}{4} \dfrac{ (1-\ihe) + 4\sin^2\theta_{W}}{Q^2 - m_{Z}^2 +i
 m_{Z} \Gamma_{Z}} & \text{for}\ B=Z \, 
\end{cases}
\ena
where the first term stands for the photon pole and the second term
stands for the $Z$ pole.
The momenta of the toponium, $\t$ and $\tb$ in the rest frame of the
toponium are given by  
\bes\bea
p^{\star\mu} &=& \sqrt{s_{\psi}} (1, \,0, \,0, \,0  )\,,
\\[3mm]
p_{1}^{\star\mu} &=& \dfrac{ \sqrt{s_{\psi}} }{2} (1+\frac{
m_{t}^{\star2} - \bar{m}_{t}^{\star2} }{s_{\psi}},
\,\beta_{t}\sin\theta^{\star}\cos\phi^{\star},
\,\beta_{t}\sin\theta^{\star}\sin\phi^{\star},
\,\beta_{t}\cos\theta^{\star}  )\,, 
\\[3mm]
p_{2}^{\star\mu} &=& \dfrac{ \sqrt{s_{\psi}} }{2} (1-\frac{
m_{t}^{\star2} - \bar{m}_{t}^{\star2} }{s_{\psi}},
\,-\beta_{t}\sin\theta^{\star}\cos\phi^{\star},
\,-\beta_{t}\sin\theta^{\star}\sin\phi^{\star},
\,-\beta_{t}\cos\theta^{\star}  )\,, 
\ena\ens
where 
\bea
\beta_{t} \aeq \sqrt{ 1 + \frac{m_{t}^{\star4}}{s_{\psi}^2} +
\frac{\bar{m}_{t}^{\star4}}{s_{\psi}^2} -\frac{ 2
m_{t}^{\star2}\bar{m}_{t}^{\star2}}{s_{\psi}^2} 
- \frac{2 m_{t}^{\star2}}{s_{\psi} }
- \frac{2 \bar{m}_{t}^{\star2}}{s_{\psi} }}.
\ena

Let us first calculate the projection operators.
Because we discuss only the S-wave toponium, there are only two kinds of
projection operators: the spin-singlet and spin-triplet projection
operators which correspond to the matrix element of operators
$\calo_{P}$ and $\calo_{V}$.
In the rest frame of the toponium we get 
\bes\bea\label{eq:projOperS}
\calp_{\th,\tbh}^{J_{\topn}=0,\ihtm}  &=& -
\frac{1}{\sqrt{2}}|\tilde{m}| e^{i
\tilde{m}\phi^{\star}}~~~~~~~~~~\;\;\,,~~~ 
(\calp_{\th,\tbh}^{J_{\topn}=0,\ihtm})^{\dag}  = -
\frac{1}{\sqrt{2}}|\tilde{m}| e^{-i \tilde{m}\phi^{\star}} 
\\[2mm]\label{eq:projOperD}
\calp_{\th,\tbh}^{J_{\topn}=1,\ihtm}  &=& 
 f(\tilde{m}, m) D^{J=1}_{\ihtm,m}(\theta^{\star}, \phi^{\star})~,~~~
(\calp_{\th,\tbh}^{J_{\topn}=1,\ihtm} )^{\dag} = 
 f^{\ast}(\tilde{m}, m) \widetilde{D}^{J=1}_{m,\ihtm}(\theta^{\star},
 \phi^{\star}) 
\ena\ens
where the helicities $m = \th -\tbh$ and $\tilde{m} = \th +\tbh$ are
defined along the top-quark momentum direction, and they are related by
the Wigner rotation to the helicity states of the toponium along its
moving direction.
The function $f(\tilde{m}, m)$ is defined as follows: 
\bea
f(\tilde{m}, m) \aeq \bigg(
\frac{1}{\sqrt{2}}\tilde{m}\sqrt{1-\beta_{t}^{2} }\,e^{i
\tilde{m}\phi^{\star}} - m \bigg)\,.
\ena 
Here we use $\widetilde{D}$ to denote the complex-conjugate-transpose of
the Wigner-D functions; see Appendix.
As we have worked in the non-relativistic approximation, the relative
momentum between top and anti-top quarks is negligible, so the
kinematical factor $\beta_{t}$ in the spin-triplet projection operator
can be neglected.

The helicity amplitudes of $\ittb h$ production are decomposed by the
type of production vertexes.
Here we use the notation $\calm_{P} (X; \th,\tbh)$ with $X=S,P,A,V,T$ to
denote their contributions, and use subscripts of $X$ to distinguish the
contributions of scalar and pseudo-scalar components of the Higgs
boson. The operators that can generate the toponium in S-wave are listed in
Table~\ref{tab:operator:swave}.

\begin{table}[htbp]
\caption{Operators which generate the toponium in S-wave}
\begin{center}
\begin{tabular}{c|c|c}
\hline\hline
{\rm Operators} & {\rm Scalar Higgs } & {\rm Pseudo-Scalar Higgs} 
\\[1mm]\hline
$\calo_{S}$  &  $~$ &  $~$
\\\hline
$ \calo_{P} $ & $\surd$ & $~$
\\\hline
$ \calo_{V} $ & $\surd$ & $~$
\\\hline
$\calo_{A}$ & $\surd$ & $~$ 
\\\hline
$\calo_{T}$ & $\surd$ & $\surd$
\\\hline\hline
\end{tabular}
\end{center}
\label{tab:operator:swave}
\end{table}%

For the scalar operator, both the scalar and pseudo-scalar components of
the Higgs boson start to contribute at P-wave, so there is no relevant
contributions. For the pseudo-scalar operator, only the scalar component of the Higgs
boson contributes, and the helicity amplitude is
\bea
\calm_{P} (P; \th,\tbh)
\aeq - \ihe {\cal G}_{\ihe}^{e} g_{h} g_{A} s \sqrt{ s_{\topn} } X_{P}
|\tilde{m}| 
e^{ - i \tilde{m}\phi^{\star}} \widetilde{D}^{J=1}_{0\ihe}(\vartheta,
\varphi)\,, 
\ena
where the kinematical factor 
\bea
X_{P} \aeq \frac{\beta}{\sqrt{2}} \sqrt{ 1 - \frac{ (m_{t}^{\star2}
-\bar{m}_{t}^{\star2})^2  }{ s_{\psi}^2 } }\,, 
\ena
which is consistent with our previous explanation in
Sec.~\ref{sec:effective-vertex} that the pseudo-scalar operator can only
generate the P-wave state of the singlet toponium and Higgs boson.
This is also true for the axial-vector operator.
The helicity amplitude is similar with $\calm_{P} (P; \th,\tbh)$, 
\bea
\calm_{P} (A; \th,\tbh)
\aeq
 \ihe {\cal G}_{\ihe}^{e} g_{h} g_{A} s \sqrt{ s_{\topn} } X_{A}
 |\tilde{m}| e^{ - i \tilde{m}\phi^{\star}
 }\widetilde{D}^{J=1}_{m\ihe}(\vartheta, \varphi)  
\ena
where the kinematical factor is
\bea
X_{A} \aeq \frac{\beta}{\sqrt{2}} \sqrt{\frac{ 4m_{t}^2 }{ s_{\topn}} }
\sqrt{ 1 - \frac{ (m_{t}^{\star2} -\bar{m}_{t}^{\star2})^2  }{
s_{\topn}^2 } }\,.
\ena
The important thing is that the contributions of pseudo-scalar and axial
vector operators are destructive.
Because only the pseudo-scalar and axial vector operators generate
the singlet toponium, therefore the total helicity amplitude for the singlet
toponium production is just the sum of these two contributions.
It is proportional to $1-\sqrt{4m_{t}^2/s_{\topn}}$, and thus negligible
near the threshold.

The triplet toponium can be produced through the vector and tensor
operators. 
The helicity amplitude for the vector operator is 
\bea
\calm_{P} (V; \th,\tbh) 
\aeq  \sum_{\ihtm'}
\ihe {\cal G}_{\ihe}^{e} g_{h} g_{V} s \sqrt{ s_{\topn} } X_{V} 
\widetilde{D}^{J=1}_{\ihtm'\ihe}(\vartheta, \varphi)
 f^{\ast}(\tilde{m},m)\widetilde{D}^{J=1}_{m \ihtm' }(\theta^{\star},
 \phi^{\star})\,.
\ena
Here the helicity $\ihtm'$ is quantized along the moving direction of
the toponium in the $e^{+}e^{-}$ rest-frame, and related to $\ihtm$ by the Wigner rotations after spin projection. The kinematical factor is 
\bea
X_{V} \aeq 2\sqrt{\frac{ 4m_{t}^2 }{ s } } \sqrt{ 1 - \frac{
(m_{t}^{\star2} -\bar{m}_{t}^{\star2})^2  }{ s_{\psi}^2 } }\,.
\ena 
This is a S-wave production, and can be represented by an effective operator
$h\topn^{\mu}B_{\mu}$, where $B=\gamma, Z$.
The contribution from the tenser operator is also of S-wave production,
and can be represented by an effective operator
$hF_{\topn\mu\nu}F_{B}^{\mu\nu}$, where $F_{B}^{\mu\nu} =
\partial^{\mu}B^\nu - \partial^{\nu}B^\mu$ is the field strength tensor.
The corresponding helicity amplitude is 
\bea
\calm_{P} (T_{S}; \th,\tbh)  \aeq \sum_{\ihtm'}
\ihe {\cal G}_{\ihe}^{e} g_{h} g_{V} s \sqrt{ s_{\topn} } X_{T_{S}} 
\widetilde{D}^{J=1}_{\ihtm'\ihe}(\vartheta, \varphi)
 f^{\ast}(\tilde{m},m)\widetilde{D}^{J=1}_{m \ihtm'}(\theta^{\star},
 \phi^{\star})\,, 
\ena
where the kinematical factor is
\bea
X_{T_{S}} \aeq 2 \sqrt{ \frac{s }{ s_{\psi} } } \bigg( 1 -
\frac{m_{h}^2}{s} \bigg)
\sqrt{ 1 - \frac{ (m_{t}^{\star2} -\bar{m}_{t}^{\star2})^2  }{
s_{\psi}^2 } }\,.
\ena 
In the above calculations we have neglected a contribution of the D-wave
production which is proportional to $\beta^2$.
Apart from the kinematical factor, the rest is completely the same as
the contribution of the vector operator $\calm_{P}(V; \th,\tbh)$.
These two contributions are constructive, and hence make the
triplet production rate dominant.
Furthermore, the pseudo-scalar component of the Higgs boson also
contributes in the S-wave toponium production via the tensor operator.
However, the overall production is of P-wave.
The corresponding effective operator can be written as
$h\widetilde{F}_{\topn\mu\nu}F_{B}^{\mu\nu}$, where
$\widetilde{F}_{B}^{\mu\nu} = 1/2
\epsilon_{\mu\nu\alpha\beta}F_{B}^{\alpha\beta}$ is the dual strength
tensor of the field $B$.
The helicity amplitude for the tensor operator is, 
\bea
\calm_{P} (T_{P}; \th,\tbh)  \aeq -\sum_{\ihtm'}
i\ihe {\cal G}_{\ihe}^{e} \epsilon_{h} g_{V}  s \sqrt{ s_{\psi} }
X_{T_{P}} \widetilde{D}^{J=1}_{\ihtm'\ihe}(\vartheta, \varphi)
 f^{\ast}(\tilde{m},m)\ihtm'\widetilde{D}^{J=1}_{m \ihtm'
 }(\theta^{\star}, \phi^{\star})\,, 
\ena
where the kinematical factor is
\bea
X_{T_{P}} \aeq 2 \beta \sqrt{ \frac{s }{ s_{\psi} } } \sqrt{ 1 - \frac{
(m_{t}^{\star2} -\bar{m}_{t}^{\star2})^2  }{ s_{\psi}^2 } },  
\ena
and $\epsilon_h=g_h\tan\xi_{ht\bar t}$.

Now we can obtain the projected helicity amplitudes.
For the pseudo-scalar and axial vector operators the projected helicity
amplitudes are similar with each other; 
\bea
\widetilde{\calm}_{P} (P; J_{\topn}=0)  
&=&  
\sqrt{2}\ihe {\cal G}_{\ihe}^{e} g_{h} g_{A} s \sqrt{ s_{\topn} } \beta
\widetilde{D}^{J=1}_{0\ihe}(\vartheta, \varphi)\,.
\\
\widetilde{\calm}_{P} (A; J_{\topn}=0)  
&= & 
-\sqrt{2}\ihe {\cal G}_{\ihe}^{e} g_{h} g_{A} s \sqrt{ s_{\topn} } \beta
\sqrt{\frac{ 4m_{t}^2 }{ s_{\topn}} }
\widetilde{D}^{J=1}_{0\ihe}(\vartheta, \varphi)\,.
\ena
Because only these two operators contribute to the singlet toponium
production, the total helicity amplitude for the singlet toponium
production is given as 
\bea
\widetilde{\calm}_{P} (\ihe; J_{\topn}=0)
=
{\cal G}_{\ihe}^{e} g_{h} g_{A} s \sqrt{ s_{\topn} } 
\beta\bigg(1-\sqrt{\frac{ 4m_{t}^2 }{ s_{\topn}} } \bigg) 
e^{ i\ihe\varphi }\sin\vartheta\,.
\ena
As expected this is the usual production helicity amplitude of two
scalar particles in P-wave.
Because it is strongly suppressed by the kinematical factor
$1-\sqrt{4m_{t}^2/s_{\topn}}$ which vanishes near the threshold.
Therefore we will neglect the singlet toponium in the following study of
spin correlations.

For the triplet toponium production, the vector and tensor operators
contributes.
Apart from the kinematical factors, the projected helicity amplitudes
have also the same structure, and proportional to the Wigner-D
function as follows: 
\bea
\widetilde{\calm}_{P} (V/T_{S}/T_{P}; J_{\psi}=1, \ihtm)  
&\propto& \widetilde{D}^{J=1}_{\ihtm\ihe}(\vartheta, \varphi)\,.
\ena
Here we have used a relation
\bea
\sum_{\th,\tbh}f^{\ast}(\tilde{m},m)f(\tilde{m},m)D^{J=1}_{\ihtm,
m}\widetilde{D}^{J=1}_{m \ihtm' } 
\aeq 2\cdot \frac{1}{2} D^{J=1}_{\ihtm, 0}\widetilde{D}^{J=1}_{0 \ihtm' }
+ \sum_{m=\pm1}D^{J=1}_{\ihtm, m}\widetilde{D}^{J=1}_{m \ihtm' }
= \delta_{\ihtm,\ihtm'}\,.
\ena
This is a usual production helicity amplitude of a vector particle.
Because the structures of the helicity amplitudes for these three
operators are the same, we can add them up directly.
After the summation, the helicity amplitudes are given by 
\bes\bea
\widetilde{\calm}_{P} (\ihe; J_{\psi}=1, \ihtm=0)  
&\propto&
-\frac{1}{\sqrt{2}}\sin\vartheta\,,
\\[2mm]
\widetilde{\calm}_{P} (\ihe; J_{\psi}=1, \ihtm=1)  
&\propto&
~~~\frac{1}{2}e^{ - i\tilde{\xi}_{h\t\tb} }(1 + \ihe\cos\vartheta)\,,
\\[2mm]
\widetilde{\calm}_{P} (\ihe; J_{\psi}=1, \ihtm=-1)  
&\propto&
~~~\frac{1}{2}e^{  i\tilde{\xi}_{h\t\tb} }(1 - \ihe\cos\vartheta)\,,
\ena\ens
where the kinematically suppressed CP phase $\tilde{\xi}_{h\t\tb}$ and
the suppression factor are defined as follows:
\bes\bea
\label{eq:tildexi}
\tan\tilde{\xi}_{h\t\tb} &=& \kappa \tan\xi_{h\t\tb}  \,,
\\[2mm]\label{eq:kappa}
\kappa &=&\beta\bigg/\bigg( 1 + \frac{ 2m_{t}\sqrt{s_{\psi}} -
m_{h}^2}{s} \bigg)\,.
\ena\ens

The production density matrix is defined as 
\bea
\rho_{P}( \ihtm, \ihtm' ) 
\aeq \sum_{\ihe=\pm1} \widetilde{\calm}_{P} (\ihe;\ihtm)
\widetilde{\calm}_{P}^{\dag} (\ihe; \ihtm') 
=  \sum_{\ihe=\pm1} \rho_{P}( \ihe; \ihtm, \ihtm' )  \,.
\ena
Inserting the helicity amplitudes we get
\bes\bea
\rho_{P}(\ihe; +, + ) &\propto& \dfrac{1}{4}(1 + \ihe\cos\vartheta)^{2}
\,, 
\\[3mm]
\rho_{P}(\ihe; 0, 0 ) &\propto& \dfrac{1}{2}\sin^2\vartheta\,,
\\[3mm]
\rho_{P}(\ihe; -, - ) &\propto& \dfrac{1}{4}(1 + \ihe\cos\vartheta)^{2}
\,, 
\\[3mm]
\rho_{P}(\ihe; +, - ) &\propto&  \dfrac{1}{4} e^{- i
2\tilde{\xi}_{h\t\tb} } \sin\vartheta^{2}
\\[3mm]
\rho_{P}(\ihe; +, 0 ) &\propto&  -\dfrac{e^{-i\tilde{\xi}_{h\t\tb}
}}{2\sqrt{2}} \sin\vartheta (1 + \ihe\cos\vartheta)
\\[3mm]
\rho_{P}(\ihe; -, 0 ) &\propto&  -\dfrac{e^{i\tilde{\xi}_{h\t\tb}
}}{2\sqrt{2}} \sin\vartheta (1 - \ihe\cos\vartheta)
\ena\ens

\subsection{Helicity amplitudes of the toponium decay}\label{sec:hela:decay} 

\begin{figure}[htb]
\begin{center}
\includegraphics[scale=0.58]{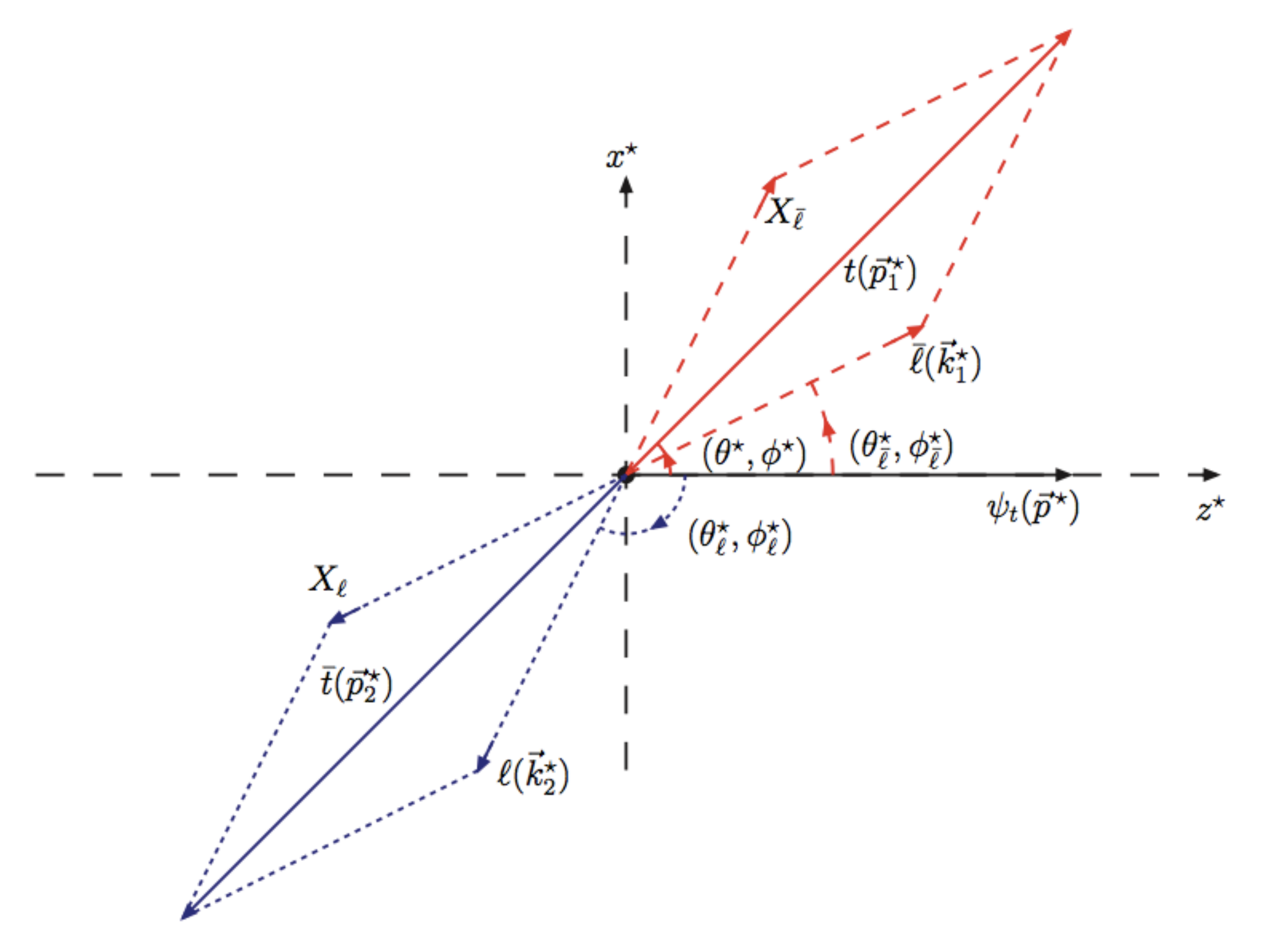} 
\caption{Definitions of the kinematical variables of top quarks and
 leptons in the toponium rest-frame.
 The $z^{\star}$ and $x^{\star}$ axes are specified by the toponium
 moving direction and the scattering plane in the laboratory frame,
 respectively.} 
\label{Fig:kin:decay}
\end{center}
\end{figure}

In this subsection we give the helicity amplitudes of the leptonic decay
of the toponium.
The kinematical variables are defined as (see also the
Fig.~\ref{Fig:kin:decay})
\bea
\widetilde{\psi}_{}( p; J_{\psi}, \ihtm ) 
\to  \t ( p_1, \th ) + \tb ( p_2, \tbh )
\to  \lb ( k_1 ) + X_{\lb} + \l ( k_2) + X_{\l} \,.
\ena
As we have mentioned the helicity amplitudes of the toponium decay are
obtained by using the spin projection of the helicity amplitudes of the
$\t\tb$ decay.
The helicity amplitudes of the $\t\tb$ decay can be separated into the
$\t$ and $\tb$ decay amplitudes as
\bea\label{eq:amp:decay:ttb}
\calm_{D}(\th, \tbh) 
\aeq \langle ( \l \nu_{\lb}\bar{b} )(\lb \nu_{\l} b)  | \calt_{D}| \t(\th)
\tb(\tbh) \rangle  
= \calm_{\t}(\th)\calm_{\tb}(\tbh)\,.
\ena
The helicity amplitudes of the $\t$ and $\tb$ decays have been known for
long time.
In the rest frame of the toponium the kinematical variables are defined
as follows:
\bes\bea
p_{1}^{\star\mu} &=& \dfrac{ \sqrt{s_{\psi}} }{2} (1+\frac{
m_{t}^{\star2} - \bar{m}_{t}^{\star2} }{s_{\psi}},
\,\beta_{t}\sin\theta^{\star}\cos\phi^{\star},
\,\beta_{t}\sin\theta^{\star}\sin\phi^{\star},
\,\beta_{t}\cos\theta^{\star}  )\,, 
\\[3mm]
p_{2}^{\star\mu} &=& \dfrac{ \sqrt{s_{\psi}} }{2} (1-\frac{
m_{t}^{\star2} - \bar{m}_{t}^{\star2} }{s_{\psi}},
\,-\beta_{t}\sin\theta^{\star}\cos\phi^{\star},
\,-\beta_{t}\sin\theta^{\star}\sin\phi^{\star},
\,-\beta_{t}\cos\theta^{\star}  )\,, 
\\[3mm]
k_{1}^{\star\mu} &=& E_{\lb} (1, \,\sin\theta^{\star}_{\lb}
\cos\phi^{\star}_{\lb} , \,\sin\theta^{\star}_{\lb}
\sin\phi^{\star}_{\lb} , \,\cos\theta^{\star}_{\lb}  )\,, 
\\[3mm]
k_{2}^{\star\mu} &=& E_{\l} (1, \,\sin\theta^{\star}_{\l}
\cos\phi^{\star}_{\l} , \,\sin\theta^{\star}_{\l} \sin\phi^{\star}_{\l}
, \,\cos\theta^{\star}_{\l}  )\,.
\ena\ens
Here and after we neglect the lepton mass.
By using the Fierzt transformation, the kinematical variables of
$(b\nu)/(\bar{b}\bar{\nu}_{\l})$ can be factorized out completely. 
Thus the anti-lepton and lepton carry all the spin informations of
$\t$ and $\tb$, respectively. 
Then the helicity amplitudes of $\t$ and $\tb$ decays can be written
as,
\bea
\calm_{\t}(\th) &=& A_{t}\sqrt{E_{t}}\sqrt{ E_{\bar{\ell}} } \,e^{i
\th\phi^{\star} }  
\nonumber \\[3mm]
&\times&
\bigg(  \cos\frac{\theta^\star}{2} \sqrt{ 1 +2\th
\cos\tilde{\theta}_{\bar{\ell}} }  
e^{i\th(\phi^{\star}_{\lb} - \phi^{\star})} 
+
2\th \sin\frac{\theta^\star}{2} \sqrt{ 1 -2\th
\cos\tilde{\theta}_{\bar{\ell}} } 
e^{-i\th(\phi^{\star}_{\lb} - \phi^{\star})}
\bigg)\,,
\ena
\bea
\calm_{\tb}(\tbh) 
&=&
A_{\tb}\sqrt{E_{t}}\sqrt{ E_{\bar{\ell}} } \, e^{i \tbh\phi^{\star} } 
\nonumber\\[3mm]
&\times&
\bigg(  \cos\frac{\theta^\star}{2} \sqrt{ 1 +2\tbh
\cos\tilde{\theta}_{\l}} 
e^{-i\tbh(\phi^{\star}_{\l} - \phi^{\star})} 
+
2\tbh \sin\frac{\theta^\star}{2} \sqrt{ 1 -2\tbh \cos\tilde{\theta}_{\l}
} 
e^{i\th(\phi^{\star}_{\l} - \phi^{\star})}
\bigg)\,,
\ena
where $A_{\t}$ and $A_{\tb}$ which are Lorentz invariant stand for the
rest of the helicity amplitude.

The $\t\tb$ decay helicity amplitudes $\calm_{D}(\th, \tbh)$ can be
obtained by using \eq{eq:amp:decay:ttb}.
In terms of $m$ and $\tilde{m}$, $\calm_{D}(\th, \tbh)$ can be
written as follows:
\bes\bea
\calm_{D}(m,\tilde{m}=0) 
&\propto&
\cos^2\frac{\theta^{\star}}{2} f_{m,-m}
-\sin^2\frac{\theta^{\star}}{2} f_{-m,m}
- \frac{1}{2}m\sin\theta^{\star}\big(  k_{m,m} - k_{-m,-m} \big)\,,
\\[3mm]
\calm_{D}(m=0,\tilde{m}) 
&\propto&
e^{i\tilde{m} \phi^{\star} }\bigg(
\cos^2\frac{\theta^{\star}}{2} k_{\tilde{m},\tilde{m}}
+\sin^2\frac{\theta^{\star}}{2} k_{-\tilde{m},-\tilde{m}}
+ \frac{\tilde{m}}{2} \sin\theta^{\star}\big(  f_{\tilde{m},-\tilde{m}}
+ f_{-\tilde{m},\tilde{m}} \big)\bigg)\,, 
\ena\ens
where the functions $f_{m,m'}$ and $k_{m,m'}$ are defined as follows:
\bea
f_{m,m'}(\phi^{\star};\theta^\star_{\lb},\phi^\star_{\lb};
\theta^\star_{\l},\phi^\star_{\l};) &=& g_{m,m'}(\theta^\star_{\lb},
\theta^\star_{\l}) e^{im(\phi^{\star}_{\lb} + \phi^{\star}_{\l})/2}
e^{im'\phi^{\star}} \,, 
\\[2mm]
k_{m,m'}(\phi^{\star};\theta^\star_{\lb},\phi^\star_{\lb};
\theta^\star_{\l},\phi^\star_{\l};) &=& g_{m,m'}(\theta^\star_{\lb},
\theta^\star_{\l}) e^{im(\phi^{\star}_{\lb} - \phi^{\star}_{\l})/2} \,, 
\\[2mm]
g_{m,m'}(\theta^{\star}_{\bar\ell},\theta^{\star}_{\ell})  &=&  
\sqrt{ 1 +m \cos\theta^{\star}_{\bar{\ell}} } 
\sqrt{ 1 +m' \cos\theta^{\star}_{\ell} }\,.
\ena

The projected helicity amplitudes can be obtained by using the
projection operators in \eq{eq:projOperS} and \eq{eq:projOperD}.
As we have explained above, the production rate of the singlet toponium
is highly suppressed near the threshold region.
Therefore we give only the decay amplitudes for the triplet toponium.
By using \eq{eq:MD}, the projected decay amplitudes for the triplet
toponium are explicitely written as
\bes\bea
\widetilde\calm_{D}(\ihtm=0) &\propto&  
\sqrt{2}\bigg(g_{1,1}(\theta^\star_{\lb}, \theta^\star_{\l})
e^{i(\phi^{\star}_{\lb} - \phi^{\star}_{\l})/2} -
g_{-1,-1}(\theta^\star_{\lb}, \theta^\star_{\l})e^{-i(\phi^{\star}_{\lb}
- \phi^{\star}_{\l})/2}\bigg)\,, 
\\[3mm]
\widetilde\calm_{D}(\ihtm=1) &\propto&  
-\, g_{1,-1}(\theta^\star_{\lb}, \theta^\star_{\l})
e^{i(\phi^{\star}_{\lb} + \phi^{\star}_{\l})/2} \,, 
\\[3mm]
\widetilde\calm_{D}(\ihtm=-1) &\propto&  
~~~g_{-1,1}(\theta^\star_{\lb}, \theta^\star_{\l})
e^{-i(\phi^{\star}_{\lb} + \phi^{\star}_{\l})/2} \,.
\ena\ens
The decay density matrix is defined as 
\bea
\rho_{D}(\ihtm,\ihtm' ) \aeq \int d\Phi( b\nu;b\bar\nu )
\calm_{D}(\ihtm)\calm_{D}^{\dag}(\ihtm')\,,
\ena
where we have integrated out the phase spaces of the bottom quarks and neutrinos. The corresponding matrix elements are
\bes\bea
\rho_{D}(0,0)
&\propto&
2\bigg(
4\cos^2\frac{\theta^{\star}_{\lb}}{2}
\cos^2\frac{\theta^{\star}_{\l}}{2} 
+ 4\sin^2\frac{\theta^{\star}_{\lb}}{2}
\sin^2\frac{\theta^{\star}_{\l}}{2} -  
2\sin\theta^{\star}_{\lb} \sin\theta^{\star}_{\l}\cos(\phi^{\star}_{\lb}
- \phi^{\star}_{\l})  
\bigg)\,,
\\[3mm]
\rho_{D}(+,+)
&\propto&
4\cos^2\frac{\theta^{\star}_{\lb}}{2}
\sin^2\frac{\theta^{\star}_{\l}}{2}\,, 
\\[3mm]
\rho_{D}(-,-)
&\propto&
4\sin^2\frac{\theta^{\star}_{\lb}}{2}
\cos^2\frac{\theta^{\star}_{\l}}{2}\,, 
\\[3mm]
\rho_{D}(0,+)
&\propto&
2\sqrt{2}\bigg( 
\sin\theta^{\star}_{\lb}\sin^2\frac{\theta^{\star}_{\l}}{2}
e^{-i\phi^{\star}_{\lb} } -
\sin\theta^{\star}_{\l}\cos^2\frac{\theta^{\star}_{\lb}}{2}
e^{-i\phi^{\star}_{\l} }
\bigg)\,,
\\[3mm]
\rho_{D}(0,-)
&\propto&
2\sqrt{2}\bigg( 
\sin\theta^{\star}_{\lb}\cos^2\frac{\theta^{\star}_{\l}}{2}
e^{i\phi^{\star}_{\lb} } -
\sin\theta^{\star}_{\l}\sin^2\frac{\theta^{\star}_{\lb}}{2}
e^{i\phi^{\star}_{\l} }
\bigg)\,,
\\[3mm]
\rho_{D}(+,-)
&\propto&
- \sin\theta^{\star}_{\lb}\sin\theta^{\star}_{\l}e^{i(\phi^{\star}_{\lb}
+ \phi^{\star}_{\l})}\,.
\ena\ens
The spin correlations occur if the imaginary part of the decay density 
matrix is non-zero.
The above results indicate that the spin correlations can appear in both
the transverse-transverse and transverse-longitudinal interferences.

\subsection{Total helicity amplitudes and CP-odd observables
  }\label{sec:hela:total} 
In this subsection we discuss the interferences among the different
helicity states of the triplet toponium.
The CP-odd observables are obtained by studying the spin correlations due
to the interferences.
As we have mentioned there are two kinds of interference:
transverse-transverse (TT) and longitudinal-transverse (LT)
interferences, which are predicted by the total density matrix,  
\bea
\rho
\aeq \sum_{\ihe=\pm1} \rho(\ihe)
= \sum_{\ihe=\pm1} \sum_{\ihtm=0,\pm1}   \sum_{\ihtm'=0,\pm1}
 \rho(\ihe; \ihtm, \ihtm' ) \,
\ena
where for convenience we have defined an intermediate density matrix as
follows:
\bea
\rho(\ihe; \ihtm, \ihtm' ) 
\aeq   
\rho_{P}(\ihe; \ihtm, \ihtm' ) \rho_{D}(\ihtm,\ihtm' )\,.
\ena
For the TT interference we have (for convenience we use $\tilde{\xi}$ to
denote the variable $\tilde{\xi}_{h\t\tb}$ for abbreviation), 
\bea
\rho(\ihe; \ihtm, -\ihtm ) &\propto& 
- \frac{1}{4}\sin\vartheta^{2}
\sin\theta^{\star}_{\lb}\sin\theta^{\star}_{\l} 
e^{i\ihtm(\phi^{\star}_{\lb} + \phi^{\star}_{\l} - 2\tilde{\xi} )}\,.
\ena
Therefore the production rate has a following non-trivial distribution with respect to
the observable $\phi^{\star}_{\lb}+\phi^{\star}_{\l}$, 
\bea\label{eq:corr:douazi}
\frac{d\sigma}{d(\phi^{\star}_{\lb}+\phi^{\star}_{\l})}
\aeq \frac{1}{2\pi}\sigma_{0}\left(1 - C_{TT}\cos(\phi^{\star}_{\lb} +
\phi^{\star}_{\l} - 2\tilde{\xi})\right)\,, 
\ena
where $\sigma_{0}$ is the total cross section, and $C_{TT}$ is the
coefficient for the $TT$ correlation.

For the LT interference we have
\bea
\rho(\ihe; 0, + ) &\propto& 
- \sin\vartheta (1 + \ihe\cos\vartheta)\bigg( 
\sin\theta^{\star}_{\lb}\sin^2\frac{\theta^{\star}_{\l}}{2}
e^{-i(\phi^{\star}_{\lb}-\tilde{\xi})} - 
\sin\theta^{\star}_{\l}\cos^2\frac{\theta^{\star}_{\lb}}{2}
e^{-i(\phi^{\star}_{\l}-\tilde{\xi}) } 
\bigg)  \,,
\\[3mm]
\rho(\ihe; 0, - ) &\propto& 
- \sin\vartheta (1 - \ihe\cos\vartheta)\bigg( 
\sin\theta^{\star}_{\lb}\cos^2\frac{\theta^{\star}_{\l}}{2}
e^{i(\phi^{\star}_{\lb}-\tilde{\xi})} -  
\sin\theta^{\star}_{\l}\sin^2\frac{\theta^{\star}_{\lb}}{2}
e^{i(\phi^{\star}_{\l}-\tilde{\xi}) } 
\bigg)  \,.
\ena
We can see that the azimuthal-angle distributions of lepton and
anti-lepton have different $\tilde{\xi}$ dependence.
The lepton momentum has a following non-trivial distribution, 
\bea\label{eq:corr:sigazi:lep}
\frac{d\sigma}{d\phi^{\star}_{\l}}
\aeq \frac{1}{2\pi}\sigma_{0}\left(1 + C_{LT}\cos(\phi^{\star}_{\l}
-\tilde{\xi})\right)\,, 
\ena
where $C_{LT}$ is the coefficient of the $LT$ correlation.
For the anti-lepton momentum, we have 
\bea\label{eq:corr:sigazi:aep}
\frac{d\sigma}{d\phi^{\star}_{\lb}}
= \frac{1}{2\pi}\sigma_{0}\left(1 - C_{LT}\cos(\phi^{\star}_{\lb} -
\tilde{\xi})\right)\,.
\ena
We can see that the correlations are different for the lepton and
anti-lepton.
For the lepton, the correlation is positive, while negative for
the anti-lepton.
On the other hand, the sign and the size of the phase shift is the same
for both the lepton and anti-lepton.
These two distributions are related with each other by the CP
transformation.
In the case of $\tilde{\xi}=0$, \ie CP is conserved, these two
correlations are symmetric under the CP transformation
$\phi^{\star}_{\lb} \to \pi - \phi^{\star}_{\l}$ and likewise for
$\phi^{\star}_{\l}$.
However, if $\tilde{\xi} \neq 0$, the distributions are asymmetric by $\tilde{\xi} \to -\tilde{\xi}$,  and therefore indicates the violation of the CP symmetry.

\section{Radiative corrections near the threshold region}\label{sec:qcd}

As we have explained in Sec.~\ref{sec:effective-vertex}, the virtual top or anti-top
quark is hugely off-shell.
According to the uncertainty principle, it can propagate only a distance
$\sim 1/(\sqrt{s}-m_{t})$ which is considerably shorter than the Coulomb
radius $r_{C} \sim 1/(\alpha_{s}m_{t})$ for the $t\bar t$ bound-state.
Therefore, near threshold production can be treated by a local source
$\delta^{4}(y_{t}-y_{\bar{t}}) j^{\mu}(Q^{2})e^{-iQ\cdot y_{t}}$.
In this approximation, the Higgs field decouples from the exact vertex
function $\langle {\rm T} h(z')\bar{t}_{i}(y_{t})
t_{j}(y_{\bar{t}})V^{\mu}(z) \rangle$ by modifying the $t\bar{t}V$
vertex function which has been examined in
Sec.~\ref{sec:effective-vertex}. 
The modified production vertexes are then in turn to affect the quantum
numbers of the generated toponia, which have been discussed in
Sec.~\ref{sec:effective-vertex}.
Here we examine how these vertexes are affected by the QCD radiative
corrections.
The corrections are described by the relativistic Salpeter-Bethe~(SB)
equation in general~\cite{SB:1951}.
For a general production vertex $\Gamma^{\mu}_{C}$ (the subscript
``$C$'' is used to distinguish possible different Dirac matrix), the SB
equation is
\bea
V^{\mu}_{C}(p, q) \aeq \Gamma^{\mu}_{C}(p, q)
+ \int\frac{d^4 k}{(2\pi)^{4}}
\calu_{\alpha\beta}(q-k)\gamma^{\alpha}S_{F}(p/2+k)V^{\mu}_{C}(p,
k)S_{F}(-p/2+k)\gamma^{\beta}\,, 
\ena
where $\calu_{\alpha\beta}(q-k)$ is the QCD potential in momentum space, and
$S_{F}$ is the Feynman propagator for fermions.
This integral equation sums over all the contributions from the relevant
ladder diagrams; see Fig.~\ref{Fig:sum}.
Here we consider only the instantaneous Coulomb-like potential, the
contributions from the transverse and rest gluons are suppressed by
powers of $\beta_{t}$. 

In the rest frame of $t\bar{t}$, the dominant contributions come from
the region where $|\vec{k}| \ll m_{t}$, and the fermionic propagators
are approximated by 
\bes\bea
S_{F}(p/2+k) &=& \frac{ i( \gamma_{+} -
\vec{k}\cdot\vec{\gamma}/(2m_{t}) ) }{ E/2 + k^{0} - \vec{k}^2/(2m_{t})
+ i\Gamma_{t}/2 } \,, 
\\[2mm]
S_{F}(-p/2+k) &=& \frac{ i( \gamma_{-} -
\vec{k}\cdot\vec{\gamma}/(2m_{t}) ) }{ E/2 - k^{0} - \vec{k}^2/(2m_{t})
+ i\Gamma_{t}/2 } \,, 
\ena\ens
where $\gamma_{\pm}=(1\pm\gamma^{0})/2$ are the non-relativistic
projection operators for fermion and anti-fermion.
Observing that the vertex function is independent of the energy $q^{0}$, 
the variable $k^{0}$ can be integrated out and we get
\bea
V^{\mu}_{C}(E, \vec{q}) \aeq \Gamma^{\mu}_{C}
- \int\frac{d^3 \vec{k}}{(2\pi)^{3}} U(\vec{q}-\vec{k})
\gamma^{0}\bigg( \gamma_{+} -
\frac{\vec{k}\cdot\vec{\gamma}}{2m_{t}}\bigg) 
\frac{V^{\mu}_{C}(E, \vec{k})}{ E - \vec{k}^2/m_{t} + i\Gamma_{t} } 
\bigg( \gamma_{-} -
\frac{\vec{k}\cdot\vec{\gamma}}{2m_{t}}\bigg)\gamma^{0}\,.
\ena

\begin{figure}[t]
\subfigure[]
{\includegraphics[scale=0.60]{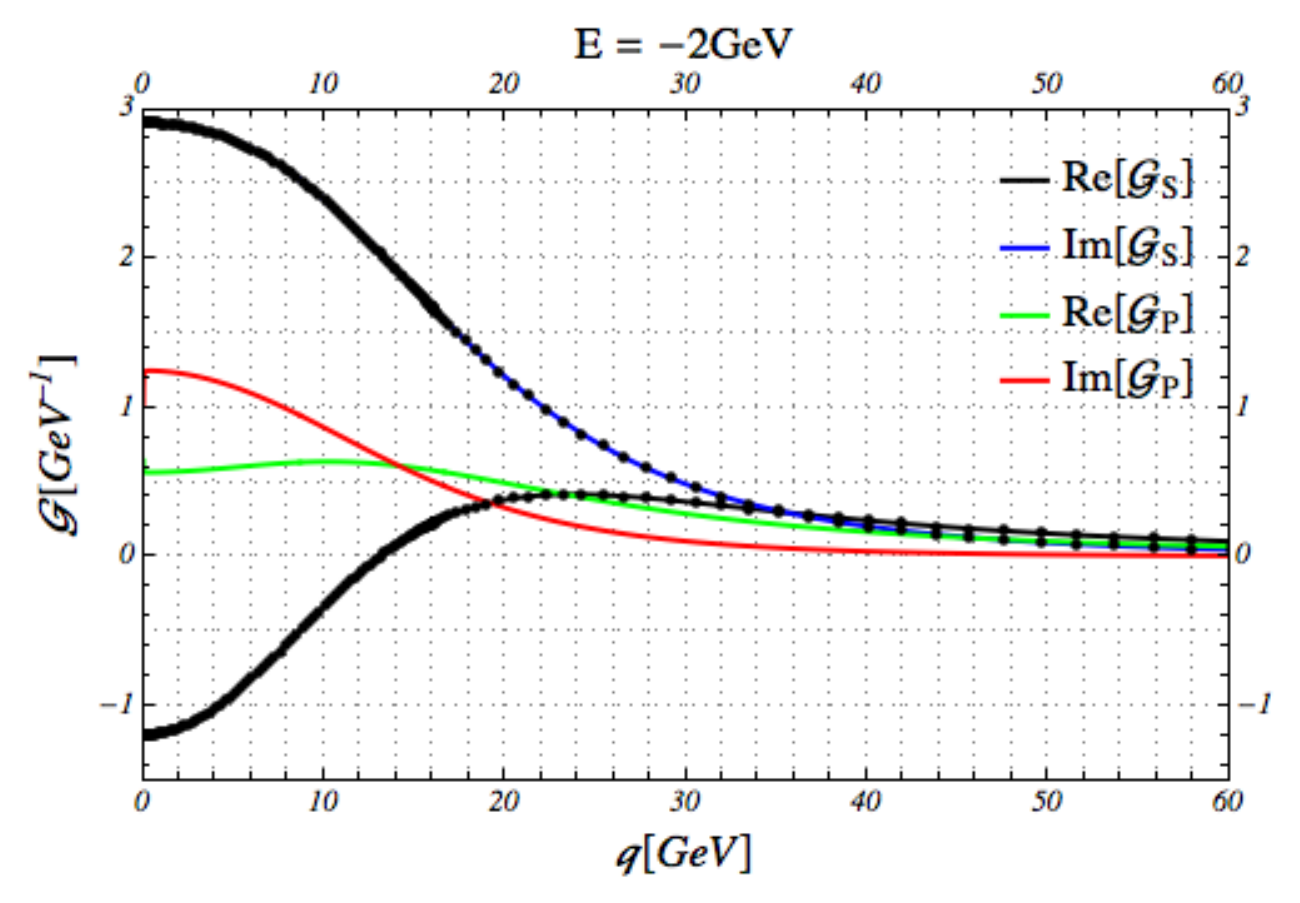}}
\hfill
\subfigure[]
{\includegraphics[scale=0.60]{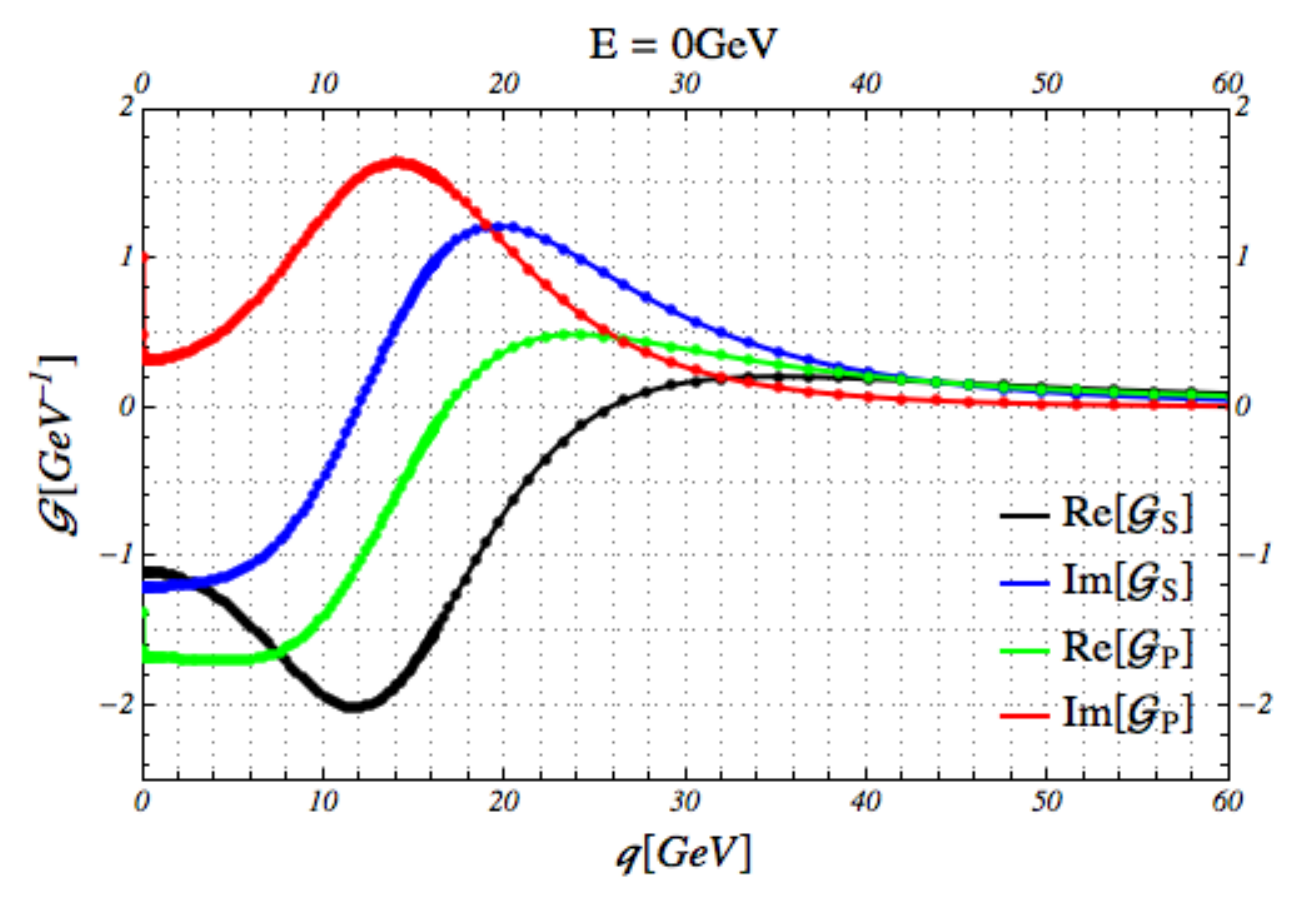}}
\hfill
\subfigure[]
{\includegraphics[scale=0.60]{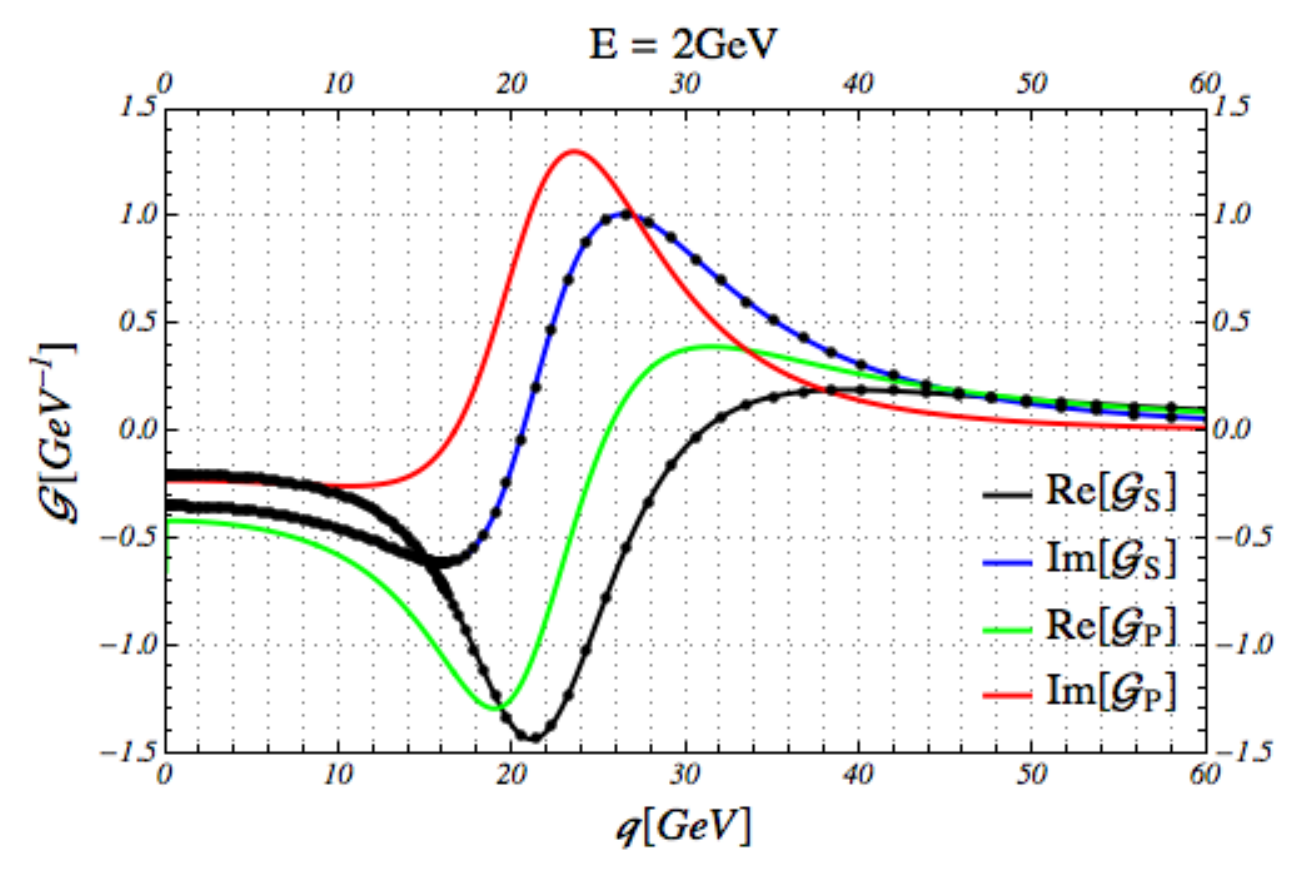}}
\hfill
\subfigure[]
{\includegraphics[scale=0.60]{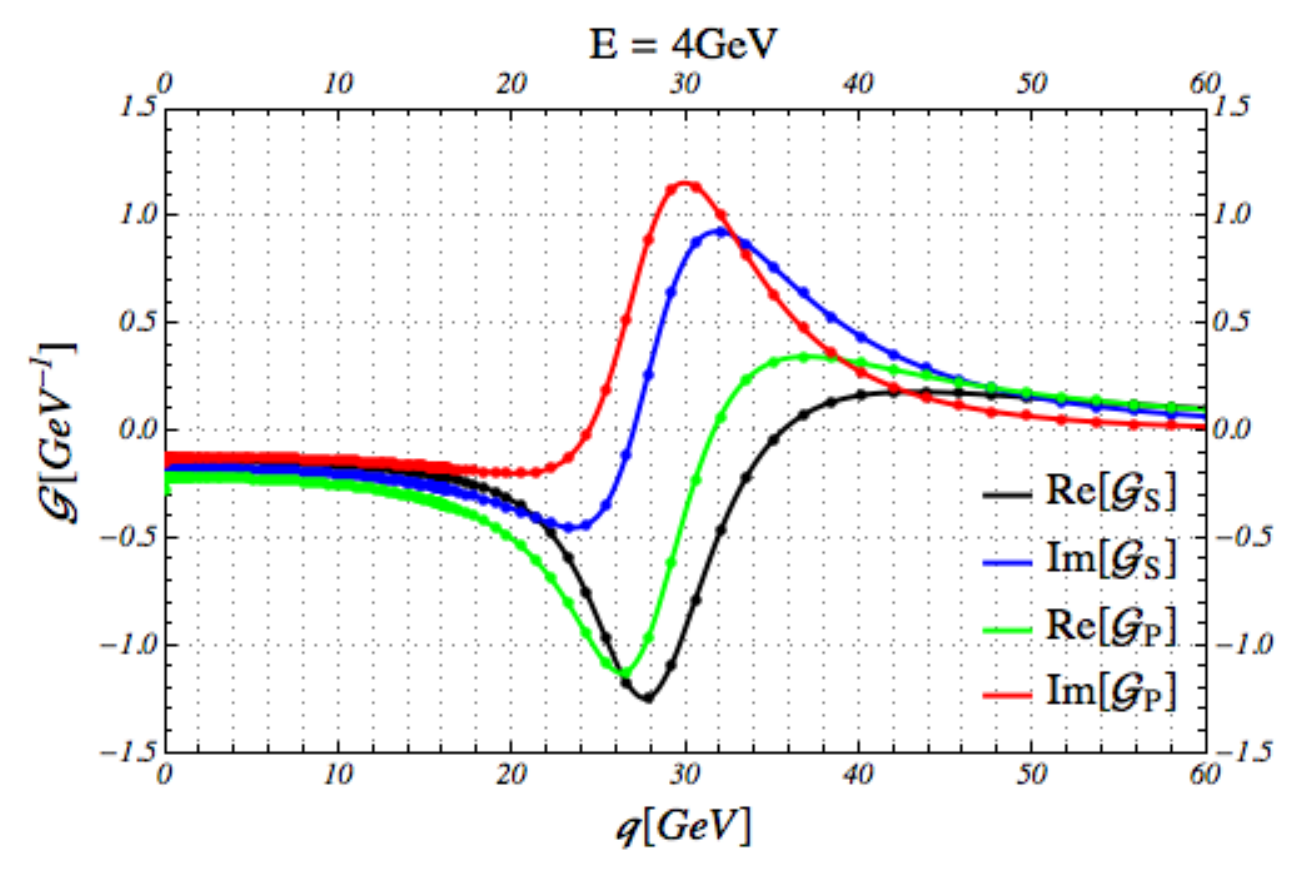}}
\caption{The Green functions as a function of $q$ for the
 four values of binding energy: (a) $E=-2~\gev$; (b) $E= 0~\gev$; (c)
 $2~\gev$; (d) $E=4~\gev$.}
\label{fig:greenf}
\end{figure}

\begin{figure}[t]
\subfigure[]
{\includegraphics[scale=0.60]{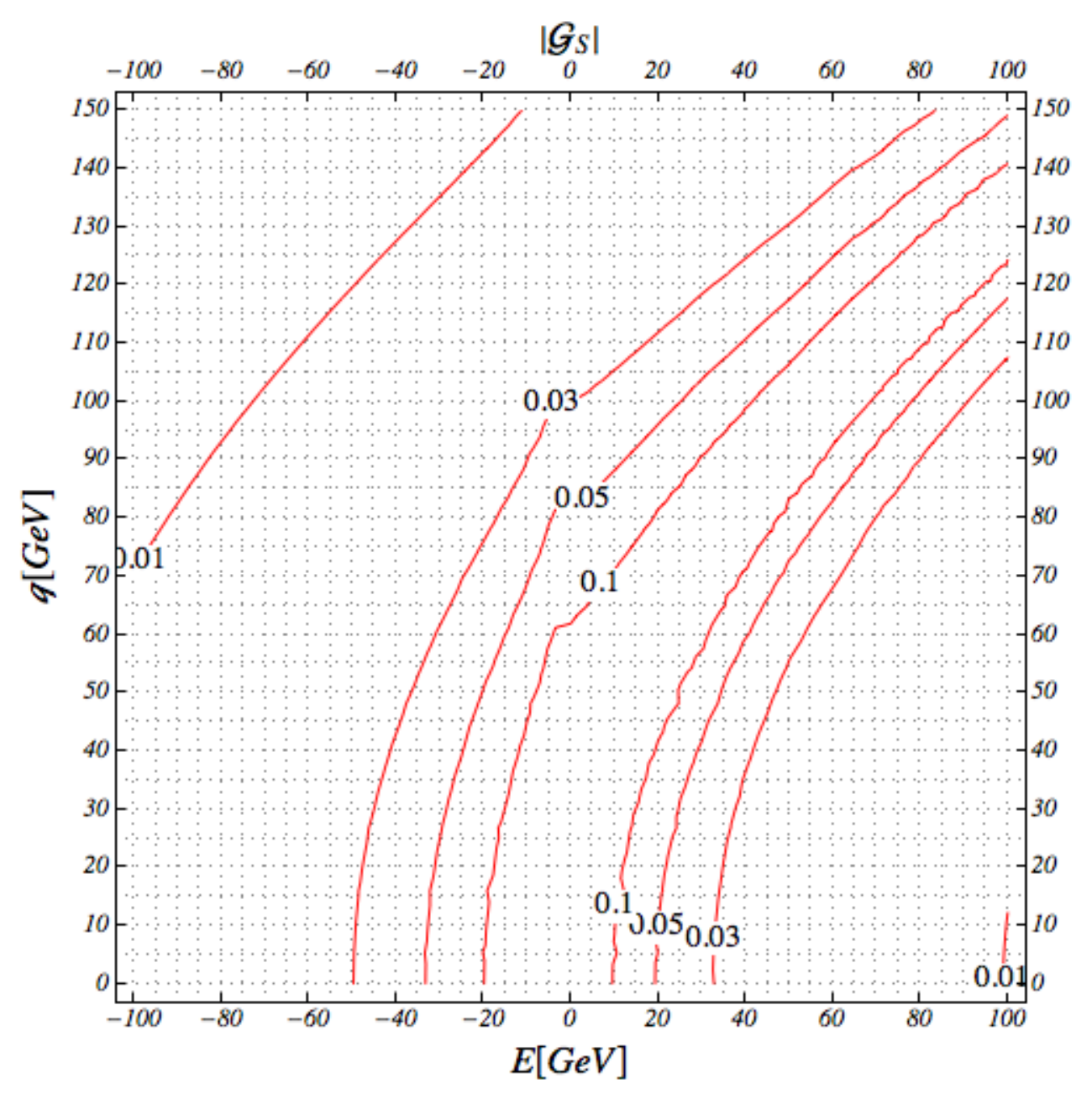}}
\hfill
\subfigure[]
{\includegraphics[scale=0.60]{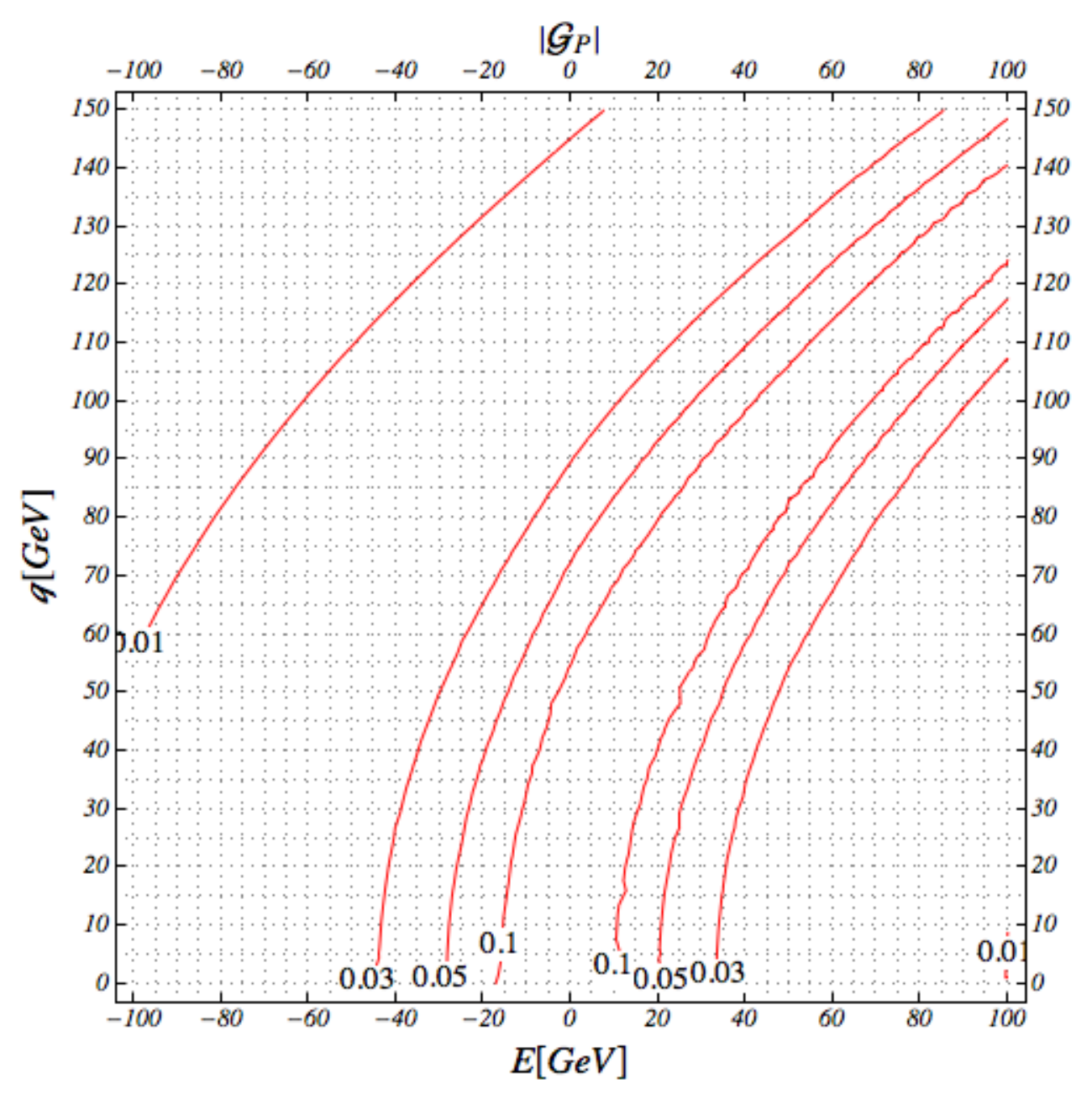}}
\caption{Contour plot of the absolute value of the Green functions for
 S-wave (a) and P-wave (b).}
\label{fig:contour:green}
\end{figure}

In our case the toponium system can be boosted by the recoil of the
Higgs boson,
therefore we express all the quantities in a Lorentz-invariant manner as
follows:
\bes\bea
E &=& \frac{1}{2} \sqrt{ (p_{1} + p_{2})^2 }= \frac{1}{2} \sqrt{ p^2
}\,, 
\\[1mm]
\gamma^{0} &=& \frac{ \pslash_{1} + \pslash_{2} }{2E} = \frac{ \pslash
}{2E} = \frac{ \pslash  }{ \sqrt{ p^2 } } \,, 
\\[1mm]
\gamma^{i}\gamma^{0} &=& \frac{1}{2}[ \gamma^{\mu}, \gamma^{0} ] = 
\frac{1}{2\sqrt{ p^2 }}[ \gamma^{\mu}, \pslash ] \equiv
\widetilde{\gamma}^{\mu}\,.
\ena\ens
The integral equation can be rewritten in a covariant form as
\bee\label{eq:con:vertex}
V^{\mu}_{C}(E, \vec{q}) = \Gamma^{\mu}_{C}
+\int\frac{d^3 \vec{k}}{(2\pi)^{3}} U(\vec{q}-\vec{k})
\bigg( \gamma_{+} - \frac{ \widetilde{\kslash} }{2m_{t}} \bigg)
\frac{V^{\mu}_{C}(E, \vec{k})}{ E - \vec{k}^2/m_{t} + i\Gamma_{t} }
\bigg( \gamma_{-} - \frac{ \widetilde{\kslash} }{2m_{t}} \bigg)\,.
\ene
We define the dressed non-relativistic projection operators for fermions
and anti-fermions as follows:
\bes\bea
\widetilde{\gamma}_{+}(\vec{q})
&=&\gamma_{+} - \frac{ \widetilde{\qslash} }{2m_{t}}
= \gamma_{+} ( 1 - \frac{ \widetilde{\qslash} }{2m_{t}})
- \gamma_{-} \frac{ \widetilde{\qslash} }{2m_{t}}\,,
\\[1mm]
\widetilde{\gamma}_{-}(\vec{q})
&=&\gamma_{-} - \frac{ \widetilde{\qslash} }{2m_{t}}
= ( 1 - \frac{ \widetilde{\qslash} }{2m_{t}} ) \gamma_{-}
-  \frac{ \widetilde{\qslash} }{2m_{t}}\gamma_{+}\,.
\ena\ens
The second terms in both $\widetilde{\gamma}_{+}(\vec{q})$ and
$\widetilde{\gamma}_{-}(\vec{q})$ involve the small component of the
Dirac spinor which are of P-wave, and therefore suppressed by an
additional factor of $\beta_{t}$.
Therefore in the following calculations we can neglect them.
In this approximation, a useful relation can be derived as follows:
\bea
\gamma_{+}\widetilde{\qslash} \aeq \widetilde{\qslash}\gamma_{-}\,.
\ena 
Multiplying $\widetilde{\gamma}_{+}(\vec{q})$ on the left-hand side and
$\widetilde{\gamma}_{-}(\vec{q})$ on the right-hand side of
\eq{eq:con:vertex}, we get 
\bea\label{eq:nonrel:vertex}
\widetilde{\gamma}_{+}(\vec{q})V^{\mu}_{C}(E, \vec{q})
\widetilde{\gamma}_{-}(\vec{q}) 
&\approx& 
\widetilde{\gamma}_{+}(\vec{q})\Gamma^{\mu}_{C}\widetilde{\gamma}_{-}(\vec{q}) 
+ \int\frac{d^3 \vec{k}}{(2\pi)^{3}} U(\vec{q}-\vec{k})
\frac{ \widetilde{\gamma}_{+}(\vec{k}) V^{\mu}_{C}(E,
\vec{k})\widetilde{\gamma}_{-}(\vec{k}) }{ E - \vec{k}^2/m_{t} +
i\Gamma_{t} }\,.
\ena
Introducing the non-relativistic reduced vertex function
\bea
\widetilde{V}^{\mu}_{C}(E, \vec{q})
\aeq\widetilde{\gamma}_{+}(\vec{q})V^{\mu}_{C}(E, \vec{q})
\widetilde{\gamma}_{-}(\vec{q})\,,\quad
\widetilde{\Gamma}^{\mu}_{C}
=\widetilde{\gamma}_{+}(\vec{q})\Gamma^{\mu}_{C}\widetilde{\gamma}_{-}(\vec{q})\,,
\ena
the integral equation \eq{eq:nonrel:vertex} reduces to
\bea
\widetilde{V}^{\mu}_{C}(E, \vec{q}) 
\aeq
\widetilde{\Gamma}^{\mu}_{C}(E, \vec{q}) 
+ \int\frac{d^3 \vec{k}}{(2\pi)^{3}} U(\vec{q}-\vec{k})
\frac{\widetilde{V}^{\mu}_{C}(E, \vec{k})  }{ E - \vec{k}^2/m_{t} +
i\Gamma_{t}}\,.
\ena
This is a formal Lippmann-Schwinger (LS) equation~\cite{LS:1950}.
Here we study only the corrections to the production vertex up to terms
linear in $\vec{q}$. 
Expanding the vertex $\widetilde{\Gamma}^{\mu}_{C}(E, \vec{q})$ by
$\vec{q}$ we have, 
\bea
\widetilde{\Gamma}^{\mu}_{C}(E, \vec{q})  
\aeq \cals^{\mu}_{C}(E )  - \calp^{\mu\nu}_{C}(E )  q_{\nu}\,,
\ena
where
\bes\bea
\cals^{\mu}_{C}(E ) &=&\widetilde{\Gamma}^{\mu}_{C}(E, \vec{q}=0)  \,,
\\[1mm]
\calp^{\mu\nu}_{C}(E ) &=& \frac{\p}{ \p q_{\nu}
}\,\widetilde{\Gamma}^{\mu}_{C}(E, \vec{q})\bigg|_{\vec{q}=0} \,, 
\ena\ens
are the S- and P-wave components, respectively.
The corrected vertex function $\widetilde{V}^{\mu}_{C}(E, \vec{q})$ can
be expanded in the same way, and we get 
\bea
\widetilde{V}^{\mu}_{C}(E, \vec{q}) 
\aeq \cals^{\mu}_{C}(E)  \calk_{S}(E, \vec{q}) +
\vec{\calp}^{\mu}_{C}(E)  \cdot \vec{q}\;\calk_{P}(E, \vec{q})\,.
\ena
The expansion coefficients satisfy following integral equations
\bes\bea
\calk_{S}(E, \vec{q}) 
&=&
1+ \int\frac{d^3 \vec{k}}{(2\pi)^{3}} U(\vec{q}-\vec{k})
\frac{ \calk_{S}(E, \vec{k})  }{ E - \vec{k}^2/m_{t} + i\Gamma_{t} }\,,
\\
\calk_{P}(E, \vec{q}) 
&=&
1+ \int\frac{d^3 \vec{k}}{(2\pi)^{3}} \frac{ \vec{q} \cdot \vec{k} }{
\vec{q}^{2} } U(\vec{q}-\vec{k}) 
\frac{ \calk_{P}(E, \vec{k})  }{ E - \vec{k}^2/m_{t} + i\Gamma_{t} }\,.
\ena\ens
These two integral equations are related to the Green function
$G(\vec{r}_{x}, \vec{r}_{x})$ which satisfies the LS equation in the
momentum space as follows: 
\bea
\bigg( E - \frac{ \vec{p}^{2} }{m_{t}} + i\Gamma_{t} \bigg)G(E;\vec{p}, \vec{r}_{y}) 
\aeq e^{i\vec{p}\cdot\vec{r}_{y}}  
+ 
\int\frac{ d^{3}\vec{k} }{(2\pi)^{3}} U(\vec{p} - \vec{k})G(E;\vec{k},
\vec{r}_{y})\,.
\ena
As we have mentioned, the local interaction approximation is excellent
in the production vertex, therefore the vertex functions are
approximated by the condition $\vec{r}_{y} = 0$.
Expanding the Green function $G(E;\vec{k}, \vec{r}_{y})$ by
$\vec{r}_{y}$ as
\bea
G(E;\vec{p}, \vec{r}_{y}) \aeq G_{S}(E;\vec{p},\vec{r}_{y}=0) +
(i\vec{r}_{y} \cdot \vec{p})\, G_{P}(E;\vec{p}, \vec{r}_{y}=0)\,, 
\ena
and the plane wave factor
$e^{i\vec{p}\cdot\vec{r}_{y}} \approx 1+i\vec{p}\cdot\vec{r}_{y}$ we obtain the
following integral equations, 
\bes\bea
\bigg( E - \frac{ \vec{p}^{2} }{m_{t}} + i\Gamma_{t}
\bigg)G_{S}(E;\vec{p}) &=& 1 
+ 
\int\frac{ d^{3}\vec{k} }{(2\pi)^{3}} U(\vec{p} -
\vec{k})G_{S}(E;\vec{k})\,, 
\\[1mm]
\bigg( E - \frac{ \vec{p}^{2} }{m_{t}} + i\Gamma_{t}
\bigg)G_{P}(E;\vec{p}) &=& 1 
+ 
\int\frac{ d^{3}\vec{k} }{(2\pi)^{3}} \frac{\vec{p}\cdot\vec{k} }{
\vec{p}^{2} } U(\vec{p} - \vec{k})G_{P}(E;\vec{k})\,.
\ena\ens
The solutions of the above equations are formally written as
\bes\bea
G_{S}(E;\vec{p}) &=& G_{0}(E;\vec{p})  + G_{0}(E;\vec{p})
\int\frac{ d^{3}\vec{k} }{(2\pi)^{3}} U(\vec{p} -
\vec{k})G_{S}(E;\vec{k})\,.
\\[1mm]
G_{P}(E;\vec{p}) &=& G_{0}(E;\vec{p}) 
+ G_{0}(E;\vec{p})
\int\frac{ d^{3}\vec{k} }{(2\pi)^{3}} \frac{\vec{p}\cdot\vec{k} }{
\vec{p}^{2} } U(\vec{p} - \vec{k})G_{P}(E;\vec{k})\,, 
\ena\ens
where $G_{0}(E;\vec{p}) $ is the Green function of a free toponium,
\bea
G_{0}(E;\vec{p}) \aeq \frac{1}{ E - \vec{p}^{2}/m_{t} + i\Gamma_{t} }\,.
\ena
These Green functions are related to the correction factors
$\calk_{S}$ and $\calk_{P}$ as follows: 
\bes\bea
G_{S}(E;\vec{p}) &=& G_{0}(E;\vec{p})\calk_{S}(E;\vec{p}),
\\
G_{P}(E;\vec{p}) &=& G_{0}(E;\vec{p})\calk_{P}(E;\vec{p}).
\ena\ens
In this study, we employ the method give in Ref.~\cite{Jezabek:1992} to
numerically solve the integral equation.
Fig.~\ref{fig:greenf} shows the S- and P-wave Green functions for
the binding energy $E\equiv\sqrt{s_{\psi_t}}-2m_t=[-2, 0, 2, 4]~\gev$.
We can see that at the ground state ($E\simeq-2~\gev$), the P-wave
contribution is suppressed.
However, the corrections on S- and P-wave are comparable for other
states.
Fig.~\ref{fig:contour:green} shows the counter lines of the absolute
values of the Green functions in the plane of the binding energy $E$ and
the relative momentum $|\vec{q}|$.  


\section{Numerical results}\label{sec:nresults}

\begin{figure}[htb]
\subfigure[]
{\includegraphics[scale=0.29]{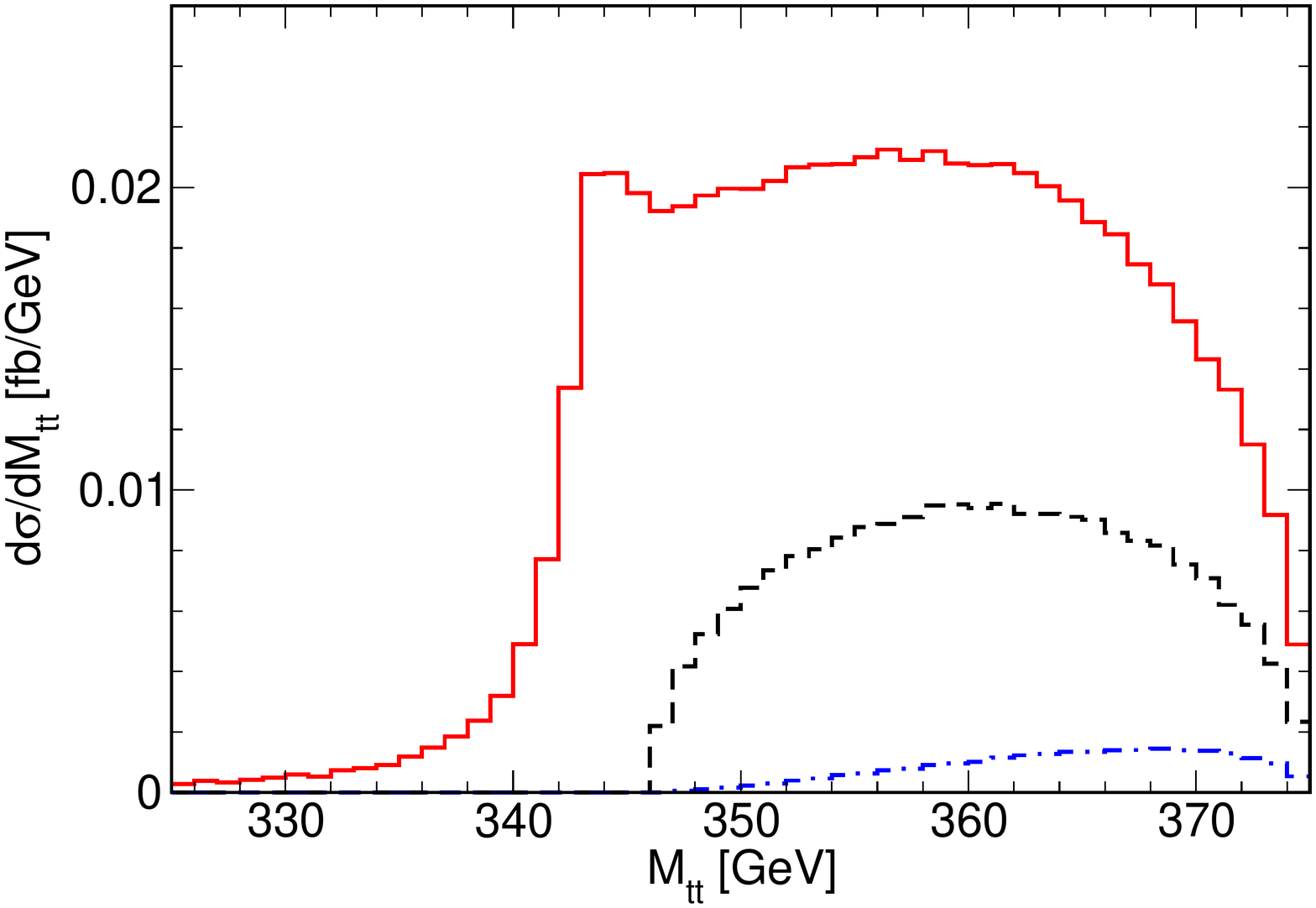}}
\subfigure[]
{\includegraphics[scale=0.29]{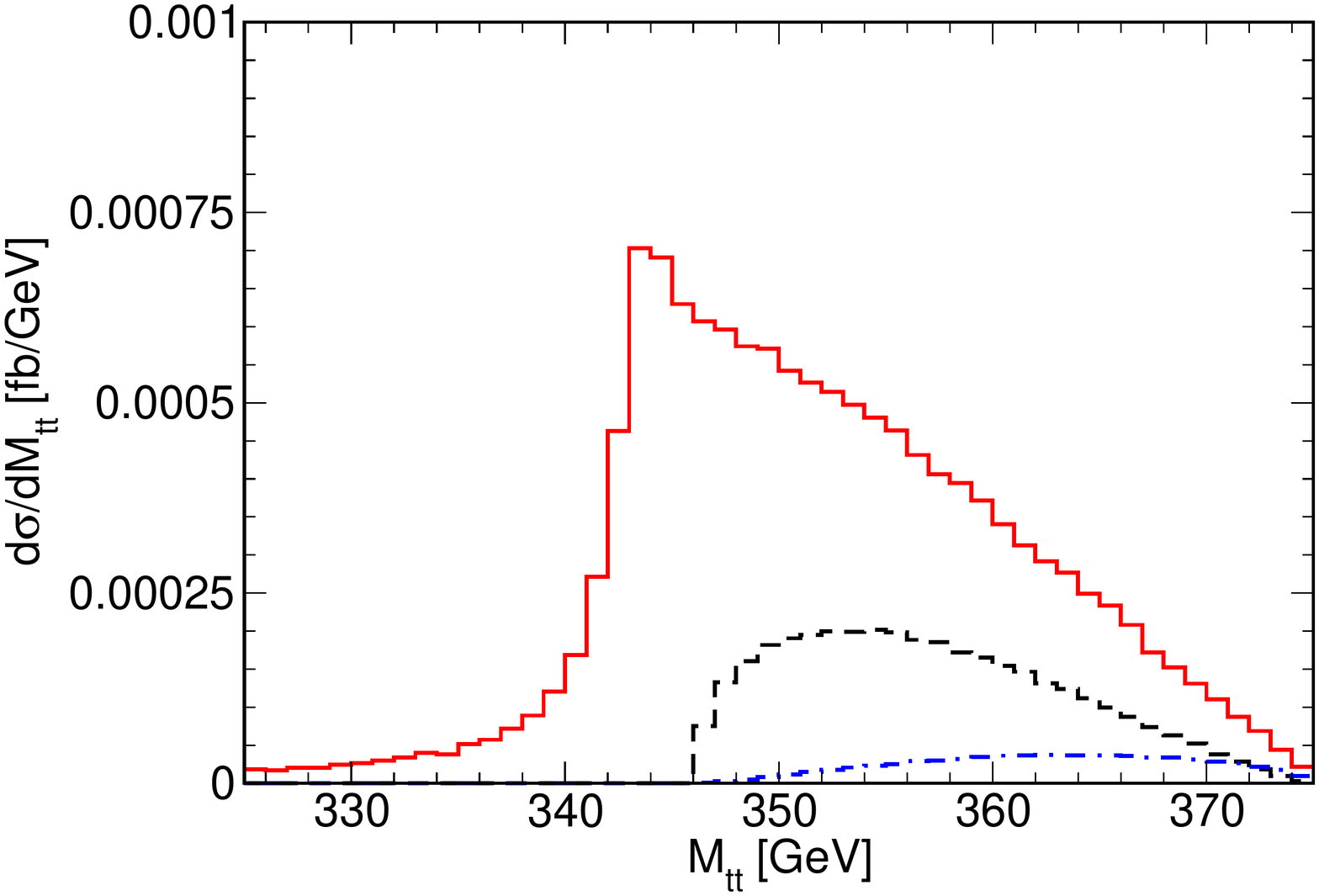}}
\caption{Production cross sections for the cases of the pure scalar
 case~(a) and pure pseudo-scalar case~(b).
 The black-dashed line is the cross section for the S-wave toponium at
 the Born level.
 The blue-dash-dotted line is the rest of the production cross
 section (which is essentially the P-wave contribution) at the Born
 level. The red-solid line is the total cross section after QCD-Coulomb
 corrections.}
\label{fig:xs:spq}
\end{figure}

Our numerical results are obtained by using MadGraph5~\cite{madgraph5}
with the HC model~\cite{mg5:model:hc} at the tree level, and then 
weighted by the QCD correction factor $\calk_{S/P}(E, \vec{q})$ at the LO. For a smooth connection to the large $M_{t\bar{t}}$ region, we follow the prescription described in Ref.~\cite{Sumino:2010bv}.
Left and right panels in Fig.~\ref{fig:xs:spq} show the production cross
sections at $\sqrt{s}=500$~GeV with respect to the invariant mass of the
$\t\tb$ system for the pure scalar and pure pseudo-scalar cases,
respectively. 
We can see that the distribution of total production cross sections have a peak below the threshold energy.
At this collision energy, the LO cross section is calculated to be
$\sigma_{\rm LO}=0.29~\fb$ for the pure scalar case (we assume that the
electron and position beams are not polarized). 
Note that when we include the diagram with a $hZZ$ vertex, the total
cross section is enhanced by about 1.7\% due to the interference with
the diagrams which contain the top-Yukawa vertex.
The QCD-Coulomb corrections give an enhancement factor of about $2.6$.
However, it has been pointed out that the NLO corrections are important
particularly in the large $\t\tb$ invariant mass
region~\cite{Farrell:2005fk,Farrell:2006}, which gives an additional
overall correction factor of about $K=0.84$~\cite{Yonamine:2011}.
In total, our prediction to the total cross section is about 0.63~fb.
We use this total cross section for the overall normalization. 
On the other hand, the NLO effects are almost uniform in the whole phase
space, therefore our LO estimation can be safely used for studying the
spin correlations.

\begin{figure}[t]
\subfigure[]
{\includegraphics[scale=0.62]{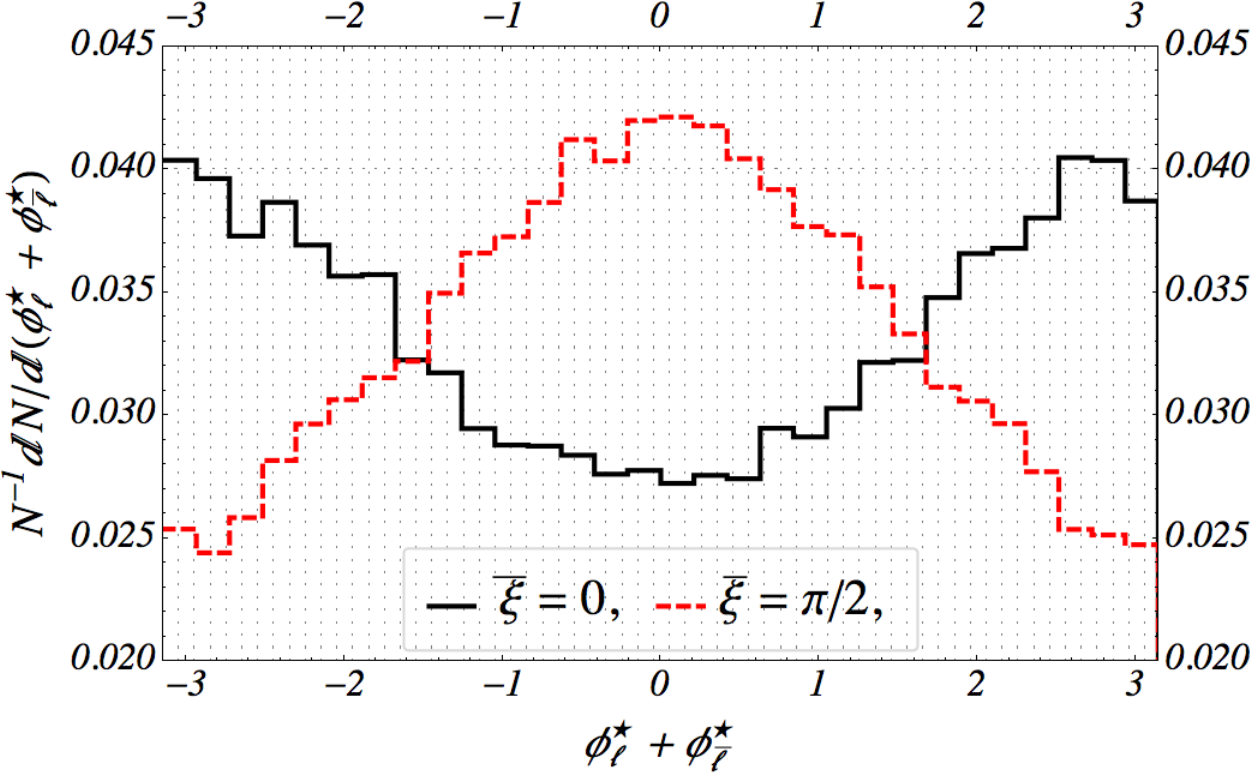}}
\hfill
\subfigure[]
{\includegraphics[scale=0.62]{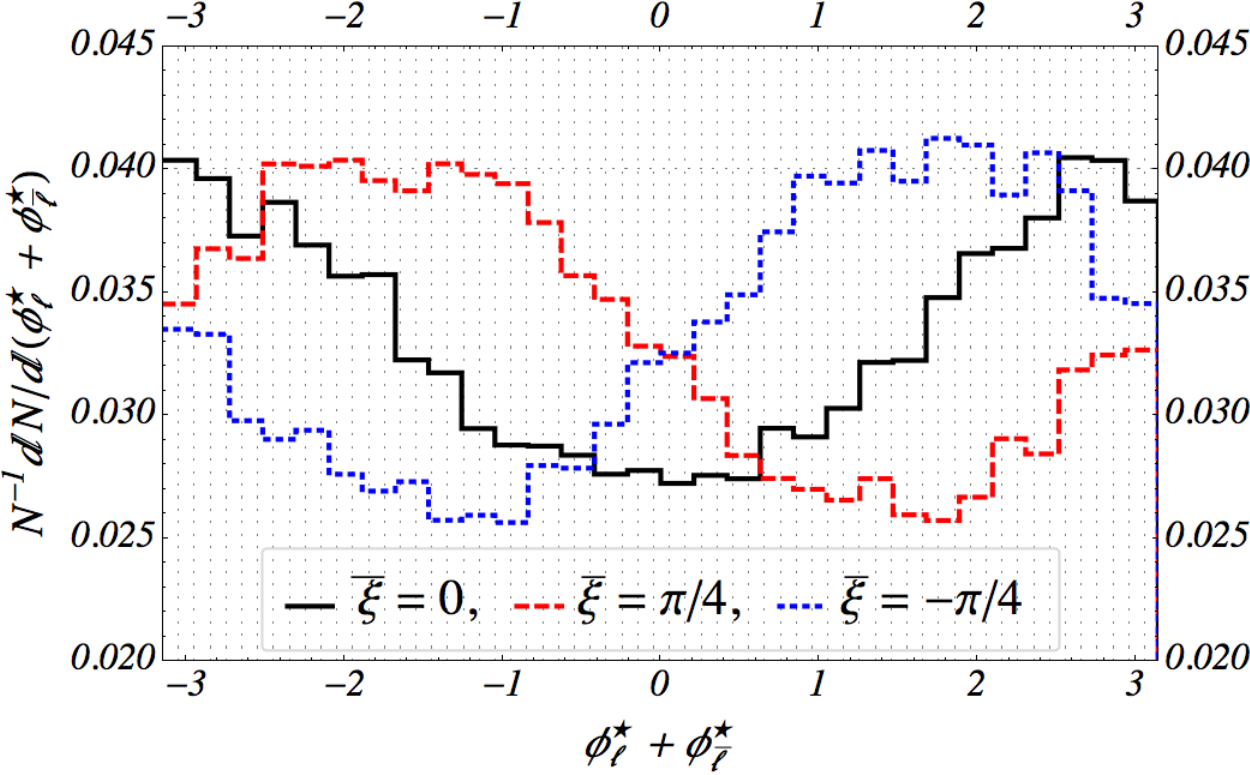}}
\caption{Azimuthal angle correlations for pure scalar and pure
 pseudo-scalar~(a) and CP-mixed cases~(b).} 
\label{fig:corr:dou}
\end{figure}

With the approximation of the S-wave dominance, we have calculated
the azimuthal angle correlations of leptons from the decays of top and
anti-top quarks.
We have shown that there are three independent CP-odd observables.
The first one is the sum of the azimuthal angles of leptons in the
toponium rest-frame, which is due to the interference among the
transverse components of the triplet toponium.
The correlation function has been given in \eq{eq:corr:douazi}.
Fig.~\ref{fig:corr:dou}~(a) shows the correlations for pure scalar Higgs
(black-solid line) and for pure pseudo-scalar Higgs (red-dashed
line).
Both are symmetric under the sign reflection of
$\phi_{\lb}^{\star}+\phi_{\l}^{\star}$, because of the CP conservation
separately. 
However the shapes are completely opposite.
In the case of scalar Higgs boson, the interference are constructive
when the sum of azimuthal angles is either $\pi$ or $-\pi$.
However it is constructive when the sum is $0$ for a pseudo-scalar
Higgs boson.
Therefore the CP violation effect is sensitive to the sign of the mixing
angle $\xi_{h\t\tb}$.
Fig.~\ref{fig:corr:dou}~(b) show three different CP-mixed cases:
$\tan\xi_{h\t\tb} = 0$~(black-solid), $\tan\xi_{h\t\tb} =
5$~(red-dashed) and $\tan\xi_{h\t\tb} = -5$~(blue-dotted).
Here in order to show the differences clearly we have chosen
$|\tan\xi_{h\t\tb}| = 5$ which means an effectively maximum mixing
because of the kinematical suppression factor $\kappa\approx 0.2$, see
\eq{eq:tildexi} and \eq{eq:kappa}.
Measuring the CP violation effects from transverse-transverse
interference requires the reconstructions of both lepton and anti-lepton
in the topponium rest-frame.
The branching ratio of top quark to leptons $(e, \mu)$ is ${\rm Br}(\t
\to \ell X) = 19\%$.
If we use the $h\to b\bar{b}$ channel to reconstruct the Higgs boson,
which has a branching ratio 56.9\%, there are 52 signal events with
$100\%$ reconstruction efficiency for the projected integrated
luminosity 4~ab$^{-1}$ at $\sqrt{s}=500~\gev$~\cite{ILC:scenarios}.
Simple estimation on the experimental sensitivity is
$\delta\tilde\xi_{h\t\t}=1.34$.
However, because of $\xi_{h\t\tb}=\tilde\xi_{h\t\tb}/\kappa$ with
$\kappa\simeq0.2$ for $\sqrt{s}=500$~GeV, the accuracy of constraining
the non-zero CP-phase may be limited at the ILC with the nominal
luminosity.
This low sensitivity comes from 1) the low total production rate, and
2) a large kinematical suppression factor $\kappa$ in
Eq.~(\ref{eq:kappa}).

\begin{figure}[t]
\subfigure[]
{\includegraphics[scale=0.62]{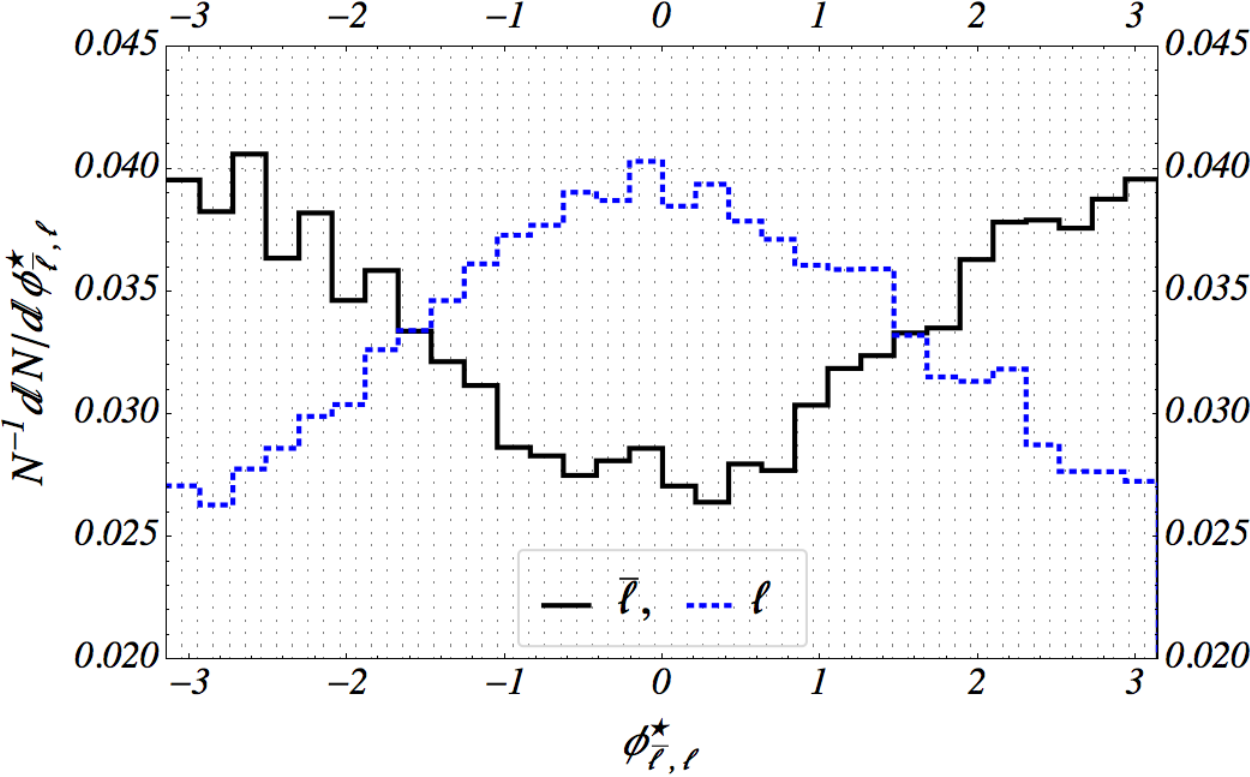}}
\hfill
\subfigure[]
{\includegraphics[scale=0.62]{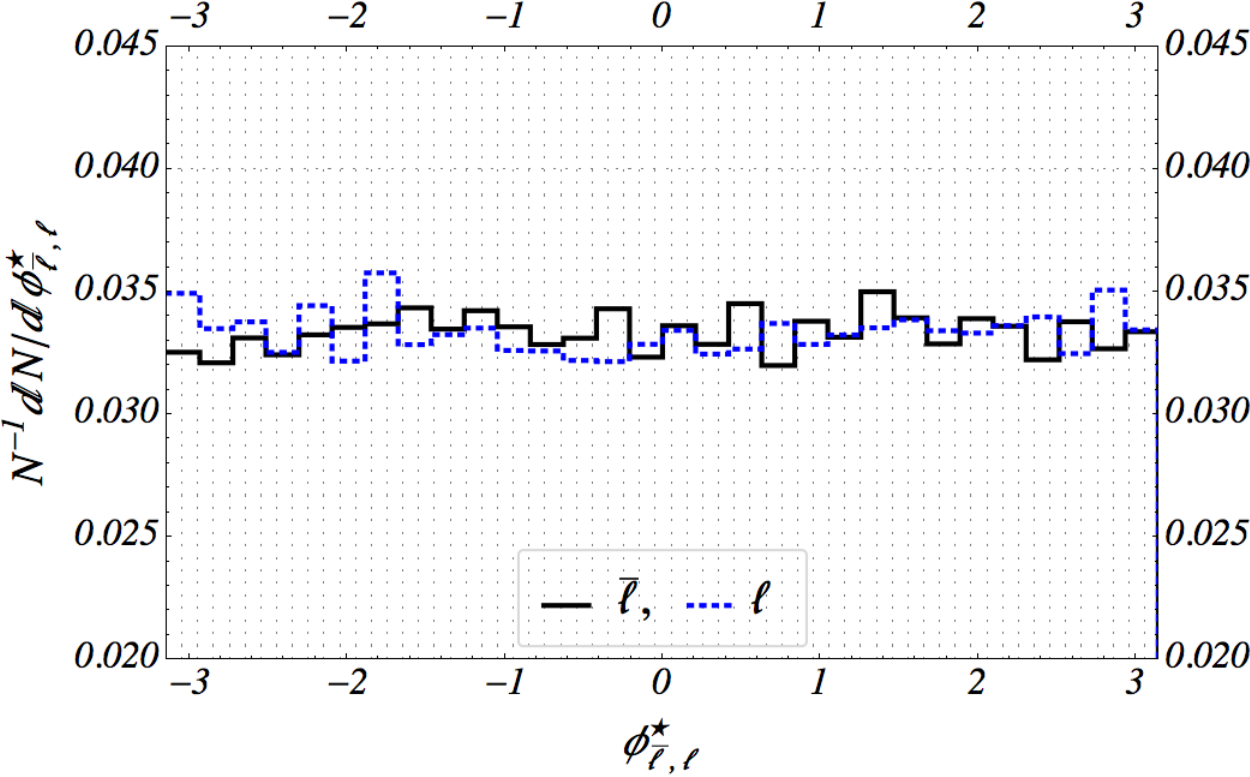}}
\hfill
\subfigure[]
{\includegraphics[scale=0.62]{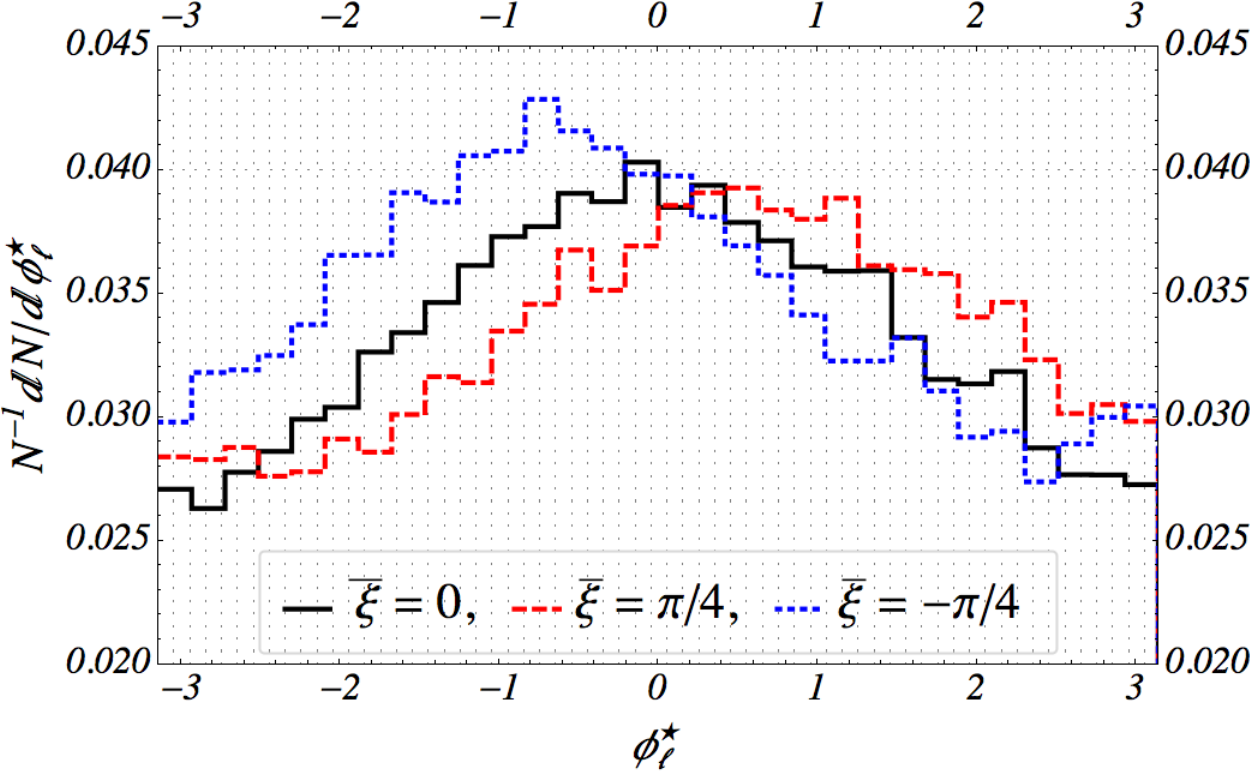}}
\hfill
\subfigure[]
{\includegraphics[scale=0.62]{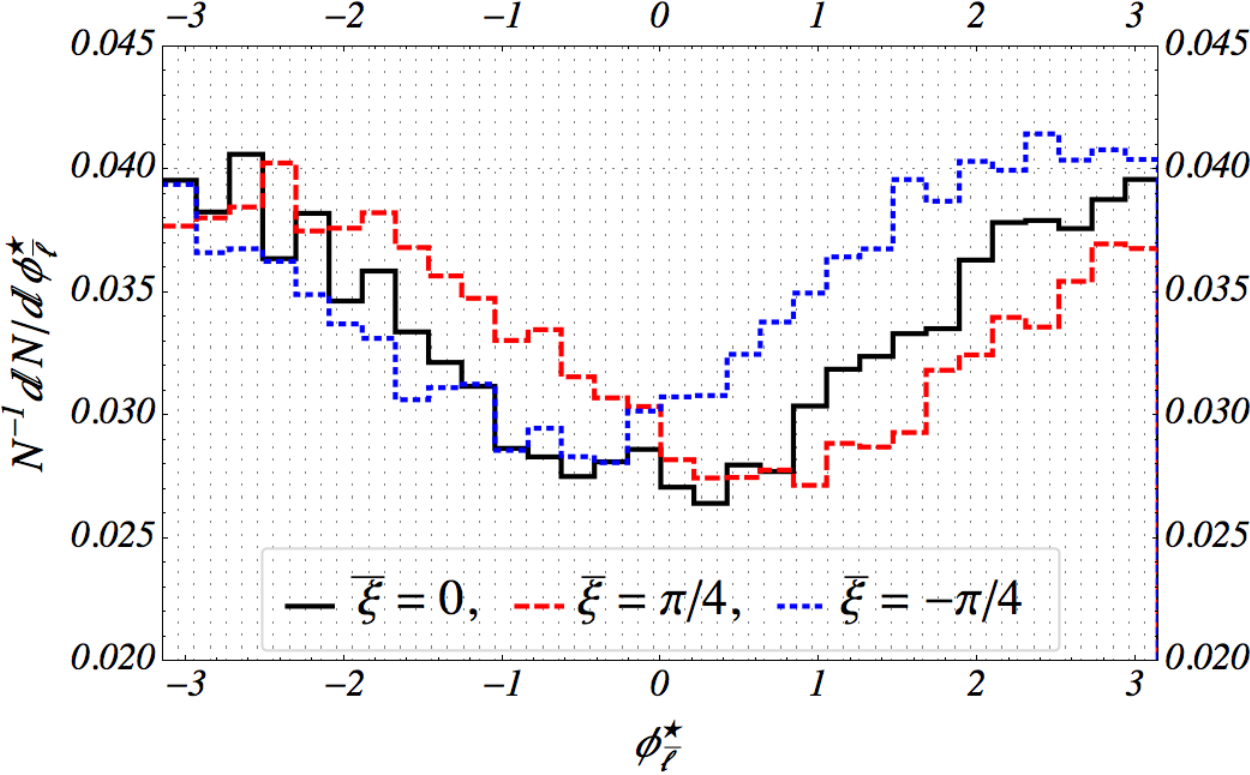}}
\caption{Azimuthal angle correlations of the lepton and anti-lepton for
 the cases of the pure scalar~(a) and the pure pseudo-scalar Higgs
 boson~(b).
 The CP-violating phase-shift is examined for $\bar\xi=0$, $\pi/4$ and
 $-\pi/4$ for the lepton~(c) and the anti-lepton distributions~(d).}
\label{fig:corr:sigazi}
\end{figure}

Apart from the interference among the transverse polarizations, there are also interferences between the longitudinally and transversely
polarized toponia which result in non-trivial azimuthal angle
distributions of the lepton and anti-lepton individually in the toponium
rest-frame.
The correlation functions are given in \eq{eq:corr:sigazi:lep} and
\eq{eq:corr:sigazi:aep}.
It is constructive at the origin ($\phi_{\l, \lb}^{\star} =0$) for
the lepton, while destructive for the anti-lepton.
For the pure scalar case, this feature is shown in
Fig.~\ref{fig:corr:sigazi}~(a).
For the pure pseudo-scalar case, because only the transversely polarized
toponium can be produced, there are no interference between
longitudinally and transversely polarized states.
Therefore the azimuthal angle distribution is flat, which is shown in
Fig.~\ref{fig:corr:sigazi}~(b).
Fig.~\ref{fig:corr:sigazi}~(c) and (d) show the interferences in the
three cases: $\tilde\xi_{h\t\tb}=0$~(black-solid),
$\tilde\xi_{h\t\tb}=\pi/4$~(red-dashed) and
$\tilde\xi_{h\t\tb}=-\pi/4$~(blue-dotted) for the lepton and
anti-lepton, respectively. 
We can see that both the lepton and anti-lepton azimuthal distributions
are sensitive to the sign of the mixing angle $\tilde{\xi}_{h\t\tb}$.
Most importantly, measuring CP violation effects through the
transverse-longitudinal interferences requires only either of the lepton 
or anti-lepton momentum being reconstructed.
For the $h\to b\bar{b}$ decay channel, there expects about 275 signal
events for the projected integrated luminosity of 4~ab$^{-1}$ at
$\sqrt{s}=500~\gev$~\cite{ILC:scenarios} (for either lepton or
anti-lepton) assuming the $100\%$ reconstruction efficiency.
Combining the lepton and anti-lepton channels we expect about 550 signal
events in total.
For this situation, the experimental sensitivity of determining
$\tilde\xi_{h\t\tb}$ is estimated to be $\delta\tilde\xi_{h\t\tb}=0.4$.
Taking into account the kinematical suppression factor of
$\kappa\simeq0.2$, the accuracy of determining $\xi_{h\t\tb}$ is
estimated to be $\delta\xi_{h\t\tb}\simeq 1.1$ for
$\xi_{h\t\tb}\simeq0$. 

\section{Discussion and Conclusion}\label{sec:conclusion}
In summary, we have studied the CP violation effects in the toponia 
productions in association with a Higgs boson at the ILC with
$\sqrt{s}=500~\gev$.
The Higgs boson can be produced by the emissions of the top or anti-top
quarks via the Yukawa interaction, or through the gauge interactions
between Higgs and vector bosons, $Z$ or $\gamma$.
The CP violation effects can appear both in the Yukawa and gauge
interactions.
However observing the effects induced by the gauge interactions is
difficult, because the CP-odd $hZZ$ interaction is induced at the loop
level while the CP-even interaction appears at the tree level.
Hence the CP asymmetry induced by the $hZZ$ coupling is
suppressed by a factor of $\alpha_{W}/(4\pi)$.
In addition, for the $e^{+}e^{-}$ production at $\sqrt{s}=500~\gev$, the
dominant contributions stem from the Higgs emissions from top quarks, but
the contributions from the gauge interactions can reach only up to a few
percent.
Therefore, CP violation effects which emerge by the top-Yukawa couplings
should be thoroughly explored.
For $\sqrt{s}=500~\gev$, the produced toponia are non-relativistic,
therefore the production of the P-wave toponium is negligible.
The eligibility of this assumption is confirmed by the numerical
calculation based on the tree-level event-generator; see
Fig.~\ref{fig:xs:spq}. 
Furthermore, the $h\t\tb$ production vertex from a virtual vector boson
$Z$ or $\gamma$ can be modeled by a contact vertex operator.
By assuming that the spins of the top and anti-top quarks are not
altered by the QCD potential, \ie the QCD potential is spin-independent,
the produced toponia spectrum are studied carefully.
In this approximation, the relevant toponia are the spin-singlet
$^{1}S_{0}$ and the spin-triplet $^{3}S_{1}$ states.
However, the production rate of the singlet toponium is found to be
highly suppressed, because it is P-wave in the toponium and Higgs boson
system. 
This observation has been checked by using the tree-level
event-generator.
Based on the careful analysis for the helicity amplitudes of the
production and decay of toponia, we propose three CP-odd observables,
namely the phase-shift in the azimuthal angle of the lepton and
anti-lepton as well as their sum. 
These observables are induced by the non-trivial correlations in the
longitudinal-transverse interferences in the azimuthal angle
distributions of leptons, and in the transverse-transverse interference
in their sum.
Compared to the up-down asymmetry examined in
Refs.~\cite{Bhupal:2008,Godbole:2007,Godbole:2011} which requires the
reconstruction of either the top- or anti-top-qaurk momenta, as well as
the small contribution from the diagram which contains $hZZ$
interactions (a few percent for $\sqrt{s}\le
1~\tev$~\cite{Bhupal:2008}), our observables do not require the
reconstruction of the momentum of the top or anti-top quark
individually, and are caused purely by the dominant $h\t\tb$
interactions.
Furthermore, all the three observables have maximum asymmetries of about
$32\%$, which are more than $6$ times larger than the maximum asymmetry
(5\%) in Refs.~\cite{Godbole:2007,Godbole:2011}.
Because the CP-odd observables for the longitudinal-transverse
interferences can be reconstructed by using only the one lepton momentum,
the number of signal events can be increased.
The experimental sensitivities for these observables are estimated
for an integrated luminosity of $L=4$~ab$^{-1}$, and found to be
$\delta\xi_{ht\bar t}\simeq 1.1$ for $\xi_{h\t\tb}=0$.
Since the sensitivity is limited mainly due to the statistical
fluctuation, it can be improved by increasing the luminosity as
projected in Ref.~\cite{ILC:scenarios}.

Compared to the current constraints on $\xi_{htt}$ by the LHC measurement, which has set  $\xi_{htt} < 0.6\pi$~\cite{Kobakhidze:2014gqa}, and further improvements by future LHC measurements, the sensitivities of our observables may be relatively low, $\delta\xi_{ht\bar t}\simeq 1.1$ for $\xi_{h\t\tb}=0$. However, our observables can be used to directly measure the CP phase, rather than to measure the overall rates. Particularly, our observables $\phi_{\ell}$ and $\phi_{\bar\ell}$ require either top or anti-top decaying to leptons, and therefore the efficiency would be enhanced dramatically.

\section{Appendix}

\subsection{Spinor wave functions in the Dirac
  representation}\label{app:dirac}
For completeness we give our conventions for the spinor wave functions
in the Dirac representation.
In the Dirac representation, Dirac matrixes are given as follows:
\bea
\gamma^{0}_{D} 
\aeq
\left(\begin{array}{cc}
1 & 0  
\\
0 & -1  
\end{array}\right)\,,\;\;\;
\vec{\gamma}_{D} 
=
\left(\begin{array}{cc}
0 & \vec{\sigma}  
\\
-\vec{\sigma} & 0  
\end{array}\right)\,.
\ena
The free solutions of the Dirac equation in the Dirac representation are 
\bea
u_{D}( \vec{p}_{1}, s ) \aeq
\left(\begin{array}{c}
\xi_{s}
\\[3mm]
\dfrac{\vec{\sigma} \cdot \vec{p}_{1} }{E+m}\xi_{s}
\end{array}\right)
\,,\;\;\;
v_{D}( \vec{p}_{2}, r ) =
\left(\begin{array}{c}
r \dfrac{\vec{\sigma} \cdot \vec{p}_{2} }{E+m}\eta_{-r}
\\[3mm]
r\eta_{-r}
\end{array}\right)\,,
\ena
where $\xi_{s}$ and $\eta_{r}$ are eigenstates of the helicity operators
$\vec{\sigma} \cdot \vec{p}_{1}/|\vec{p}_{1}|$ and $\vec{\sigma} \cdot
\vec{p}_{2}/|\vec{p}_{2}|$, respectively.
For completeness we also give the helicity eigenstates as follows:
\bea
\xi_{+} \aeq
\left(\begin{array}{c}
\cos(\theta/2)
\\
e^{i\phi}\sin(\theta/2)
\end{array}
\right)
\,,\;\;\;
\xi_{-} = 
\left(\begin{array}{c}
-e^{-i\phi}\sin(\theta/2)
\\
\cos(\theta/2)
\end{array}
\right)\,.
\ena

The spinor wave functions and the Dirac gamma matrices in the Dirac
representation are related to the ones in the chiral representation by
the following unitary transformation: 
\bea
\psi_{D} \aeq  U_{D} \psi U_{D}^{-1}
\,,\;\;\;
\gamma^{\mu}_{D} = U_{D} \gamma_{C}^{\mu}U_{D}^{-1}
\,,\;\;\;
U_{D} = \frac{1}{\sqrt[]{2}}
\left(\begin{array}{cc}
1 & 1
\\
-1 & 1
\end{array}\right)\,.
\ena

\subsection{Vector wave functions and Wigner-D functions}\label{subsec:vectorWave}

The helicity wave functions polarized along the direction $\vec{n} =
(\sin\theta\cos\phi,\, \sin\theta\sin\phi,\, \cos\theta)$ for vector
particles in the rest frame are defined as follows:
\bea\label{eq:vectorWaveT}
\epsilon(\vec{n}, \lambda=\pm1)
&=& \dfrac{1}{\sqrt{2}} (0,\, -\lambda\cos\theta\cos\phi + i\sin\phi,\,
-\lambda\cos\theta\sin\phi - i \cos\phi,\, \lambda\sin\theta )\,, 
\\[3mm]\label{eq:vectorWaveL}
\epsilon(\vec{n}, \lambda=0)
&=&
(0,\, \sin\theta\cos\phi,\, \sin\theta\sin\phi,\, \cos\theta)
\ena

The Wigner-D function for spin-1 particle is defined as follows:
\bea
\vec{\epsilon}( \vec{n}, \lambda' ) \aeq
\sum_{\lambda=0,\pm1} D^{J=1}_{\lambda\lambda'}(\theta, \phi)
\,\vec{\epsilon}( \vec{0}, \lambda )
\ena
and it's inverse
\bea
\vec{\epsilon}( \vec{0}, \lambda ) \aeq 
\sum_{\lambda''=0,\pm1} \widetilde{D}^{J=1}_{\lambda''\lambda}(\theta,
\phi) 
\,\vec{\epsilon}( \vec{n}, \lambda'' ) 
\ena
and following relation holds
\bea
\widetilde{D}^{J=1}_{\lambda'\lambda}(\theta, \phi) \aeq
(D^{J=1}_{\lambda\lambda'}(\theta, \phi))^{\ast} 
\ena
Based on these definitions we also have
\bea
\vec{\epsilon}^{\;\ast}( \vec{n}, \lambda' ) \cdot \vec{\epsilon}(
\vec{0}, \lambda ) 
&=& 
\widetilde{D}^{J=1}_{\lambda'\lambda}(\theta, \phi)
\\[2mm]
\vec{\epsilon}^{\;\ast}( \vec{0}, \lambda ) \cdot \vec{\epsilon}(
\vec{n}, \lambda' )  
&=&
D^{J=1}_{\lambda\lambda'}(\theta, \phi)
\ena

\section*{Acknowledgments}
K.H.\ is supported by the William F.\ Vilas Trust Estate, and by the
U.S.\ Department of Energy under the contract DE-FG02-95ER40896.
K.M.\ is supported by the China Scholarship Council and the Hanjiang
Scholar Project of Shaanxi University of Technology.
The work of H.Y.\ was supported by JSPS KAKENHI Grant Number 15K17642.


\end{document}